\documentclass{article}
\pdfoutput=1

\usepackage{tcolorbox}
\usepackage{wrapfig}
\usepackage{enumitem}
\usepackage{amssymb,amsbsy,amsmath,amsfonts,amssymb,amscd,bm}
\usepackage{calrsfs}
\usepackage[makeroom]{cancel}
\usepackage{array}
\DeclareMathAlphabet{\pazocal}{OMS}{zplm}{m}{n}
\newcommand{\La}{\mathcal{L}}

\setlist{nolistsep}
\usepackage{color}

\usepackage{mathtools}
\usepackage{tikz}
\makeatletter
\newcommand\mathcircled[1]{%
  \mathpalette\@mathcircled{#1}%
}

\newcommand{\e}{\varepsilon}
\newcommand{\ed}{\dt\varepsilon}
\newcommand{\edd}{\dtt\varepsilon}
\newcommand{\s}{\sigma}
\newcommand{\sd}{\dt\sigma}
\newcommand{\sdd}{\dtt\sigma}
\newcommand {\dt} {\partial_{t}}
\newcommand {\dx} {\partial_{x}}
\newcommand {\dtt} {\partial_{tt}^2}
\newcommand {\dxx} {\partial_{xx}^2}

\newcommand {\bn} {\bar{n}}
\newcommand {\br} {\bar{\rho}}
\newcommand {\bu} {\bar{u}}

\title{Mechanical models of pattern and form in biological tissues: the role of stress-strain constitutive equations  
}

\author{Chiara Villa \and
Mark A. J. Chaplain  \and
Alf Gerisch \and
        Tommaso Lorenzi                   
}

\author{Chiara Villa\thanks{School of Mathematics and Statistics, University of St Andrews, UK 
  (cv23@st-andrews.ac.uk)}
\and
Mark A. J. Chaplain\thanks{School of Mathematics and Statistics, University of St Andrews, UK 
  (majc@st-andrews.ac.uk)}
\and
Alf Gerisch\thanks{Department of Mathematics, Technische Universität Darmstadt, Germany
 (gerisch@mathematik.tu-darmstadt.de) }
 \and
Tommaso Lorenzi\thanks{Department of Mathematical Sciences ``G. L. Lagrange'', Dipartimento di Eccellenza 2018-2022, Politecnico di Torino, 10129 Torino, Italy
  (tommaso.lorenzi@polito.it)}
}

\begin{document}
\date{}
\maketitle

\begin{abstract}
Mechanical and mechanochemical models of pattern formation in biological tissues have been used to study a variety of biomedical systems, particularly in developmental biology, and describe the physical interactions between cells and their local surroundings. These models in their original form consist of a balance equation for the cell density, a balance equation for the density of the extracellular matrix (ECM), and a force-balance equation describing the mechanical equilibrium of the cell-ECM system. Under the assumption that the cell-ECM system can be regarded as an isotropic linear viscoelastic material, the force-balance equation is often defined using the Kelvin-Voigt model of linear viscoelasticity to represent the stress-strain relation of the ECM. However, due to the multifaceted bio-physical nature of the ECM constituents, there are rheological aspects that cannot be effectively captured by this model and, therefore, depending on the pattern formation process and the type of biological tissue considered, other constitutive models of linear viscoelasticity may be better suited. In this paper, we systematically assess the pattern formation potential of different stress-strain constitutive equations for the ECM within a mechanical model of pattern formation in biological tissues. The results obtained through linear stability analysis and the dispersion relations derived therefrom support the idea that {fluid-like constitutive models}, such as the Maxwell model and the {Jeffrey} model, have a pattern formation potential much higher than {solid-like models}, such as the Kelvin-Voigt model and the standard linear solid model. This is confirmed by the results of numerical simulations, which demonstrate that, all else being equal, spatial patterns emerge in the case where the Maxwell model is used to represent the stress-strain relation of the ECM, while no patterns are observed when the Kelvin-Voigt model is employed. Our findings suggest that further empirical work is required to acquire detailed quantitative information on the mechanical properties of components of the ECM in different biological tissues in order to furnish mechanical and mechanochemical models of pattern formation with stress-strain constitutive equations for the ECM that provide a more faithful representation of the underlying tissue rheology.
\end{abstract}

\section{Introduction}
\label{sec:intro}
Pattern formation resulting from spatial organisation of cells is at the basis of a broad spectrum of physiological and pathological processes in living tissues~\cite{jernvall2003mechanisms}. {While the first formal exploration of pattern and form from a mathematical (strictly speaking, geometrical) perspective goes back over a century to D'Arcy Thompson's ``{\it On Growth and Form}'' \cite{darcyt1917}}, the {modern} development of mathematical models for this biological phenomenon started halfway through the twentieth century to elucidate the mechanisms that underly morphogenesis and embryogenesis~\cite{maini2005morphogenesis}. Since then, a number of mathematical models for the formation of cellular patterns have been developed~\cite{urdy2012evolution}. Amongst these, particular attention has been given to reaction-diffusion models and mechanochemical models of pattern formation~\cite{murray2001mathematical}.

Reaction-diffusion models of pattern formation, first proposed by Turing in his seminal 1952 paper~\cite{turingchemical} and then further developed by Gierer and Meinhardt~\cite{gierermeinhardt1972,meinhardt1982}, apply to scenarios in which the heterogeneous spatial distribution of some chemicals (\emph{i.e.} morphogens) acts as a template ({\it i.e.} a pre-pattern) according to which cells organise and arrange themselves in different sorts of spatial patterns. These models are formulated as coupled systems of reaction-diffusion equations for the space-time dynamics of the concentrations of two morphogens, with different reaction kinetics depending on the biological problem at stake. Such systems exhibit diffusion-driven instability whereby homogenous steady states are driven unstable by diffusion, resulting in the formation of pre-patterns, provided that the diffusion rate of one of the morphogens is sufficiently higher than the other~\cite{maini1997spatial,maini2019turing,maini2012turing,murray1981}. 

On the other hand, mechanochemical models of pattern formation, first proposed by Murray, Oster and coauthors in the 1980s~\cite{murray1984cell,murray1984generation,murray1983mechanical,oster1983mechanical}, describe spatial organisation of cells driven by the mechanochemical interaction between cells and the extracellular matrix (ECM) -- \emph{i.e.} the substratum composed of collagen fibers and various macromolecules, partly produced by the cells themselves, in which cells are embedded~\cite{harris1984tissue,harris1981fibroblast}. These models in their original form consist of systems of partial differential equations (PDEs) comprising a balance equation for the cell density, a balance equation for the ECM density, and a force-balance equation describing the mechanical equilibrium of the cell-ECM system~\cite{murray1989pattern,murray1988mechanochemical}. When chemical processes are neglected, these models reduce to mechanical models of pattern formation~\cite{byrne1996importance,murray1989pattern,murray1988mechanochemical}.

{While reaction-diffusion models well explain the emergence and characteristics of patterns arising during chemical reactions~\cite{castets1990experimental,maini1997spatial,maini2019turing}, as well as pigmentation patterns found on shells~\cite{meinhardt2009algorithmic} or animal coatings~\cite{kondo1995reaction,murray2001mathematical}, various observations seem to suggest they may not always be the most suited models to study morphogenic pattern formation~\cite{bard1974well,brinkmann2018post,maini2019turing}. 
For instance, experiments up to this day seem to fail in the identification of appropriate morphogens and overall molecular interactions predicted by Turing models in order for \textit{de novo} patterns to emerge may be too complex. In addition, unrealistic parameter values would be required in order to reproduce experimentally observable patterns and the models appear to be too sensitive to parameter changes, hence lacking the robustness required to capture precise patterns.
These considerations indicate that other mechanisms, driven for instance by significant 
mechanical forces, should be considered since solely chemical interactions may not suffice in explaining the emergence of patterns during morphogenesis. Hence mechanochemical models may be better suited. Interestingly, this need to change modelling framework sometimes arises within the same biological application as time progresses. For instance, supracellular organisation in the early stages of embryonic development closely follows morphogenic chemical patterns, but further tissue-level organisation requires additional cooperation of osmotic pressures and mechanical forces~\cite{petrolli2019confinement}. Similarly, pattern formation during vasculogenesis is generally divided into an early stage highly driven by cell migration following chemical cues, and a later one dominated by mechanical interactions between the cells and the ECM~\cite{ambrosi2005review,scianna2013review,tosin2006mechanics}. Finally, purely mechanical models are a useful tool for studying the isolated role of mechanical forces and can capture observed phenomena without the inclusion of chemical cues~\cite{petrolli2019confinement,serra2012mechanical,tlili2018collective}.}

Over the years, mechanochemical and mechanical models of pattern formation in biological tissues have been used to study a variety of biomedical problems, including morphogenesis and embryogenesis~\cite{brinkmann2018post,cruywagen1992tissue,maini1988nonlinear,murray1986new,murray1988mechanochemical,murray1984cell,murray1984generation,murray1983mechanical,oster1983mechanical,perelson1986nonlinear}, angiogenesis and vasculogenesis~\cite{manoussaki2003mechanochemical,scianna2013review,tranqui2000mechanical}, cytoskeleton reorganisation~\cite{alonso2017mechanochemical,lewis1991analysis}, wound healing and contraction~\cite{javierre2009numerical,maini2002mathematical,olsen1995mechanochemical,tranquillo1992continuum}, and stretch marks~\cite{gilmore2012mechanochemical}. These models have also been used to estimate the values of cell mechanical parameters, with a particular focus on cell traction forces~\cite{barocas1995fibroblast,barocas1994biphasic,bentil1991pattern,ferrenq1997modelling,moon1993fibroblast,perelson1986nonlinear}. The roles that  different biological processes play in the formation of cellular patterns can be disentangled via linear stability analysis (LSA) of the homogenous steady states of the model equations -- \emph{i.e.} investigating what parameters of the model, and thus what biological processes, can drive homogenous steady states unstable and promote the emergence of cell spatial organisation. Further insight into certain aspects of pattern formation in biological tissues can also be provided by nonlinear stability analysis of the homogenous steady states~\cite{cruywagen1992tissue,lewis1991analysis,maini1988nonlinear}. 

These models usually rely on the assumption that the cell-ECM system can be regarded as an isotropic linear viscoelastic material. This is clearly a simplification due to the non-linear viscoelasticity and anisotropy of soft tissues~\cite{bischoff2004rheological,huang2005quasi,liu2000viscoelastic,nasseri2002viscoelastic,snedeker2005strain,valtorta2005dynamic,verdier2003rheological}, a simplification that various rheological tests conducted on biological tissues have nonetheless shown to be justified in the regime of small strains~\cite{bilston1997linear,liu2000viscoelastic,nasseri2002viscoelastic,valtorta2005dynamic}, which is the one usually of interest in the applications of such models. Under this assumption, the force-balance equation for the cell-ECM system is often defined using the Kelvin-Voigt model of linear viscoelasticity to represent the stress-strain relation of the ECM~\cite{byrne1996importance,murray1988mechanochemical,oster1983mechanical}. However, due to the multifaceted bio-physical nature of the ECM constituents, there are rheological aspects that cannot be effectively captured by the Kelvin-Voigt model and, therefore, depending on the pattern formation process and the type of biological tissue considered, other constitutive models of linear viscoelasticity may be better suited~\cite{barocas1994biphasic}. In this regard, \cite{byrne1996importance} demonstrated that, {\it ceteris paribus}, using the Maxwell model of linear viscoelasticity to describe the stress-strain relation of the ECM in place of the Kelvin-Voigt model can lead to different dispersion relations with a higher pattern formation potential. This suggests that a more thorough investigation of the capability of different stress-strain constitutive equations of producing spatial patterns is required. 

With this aim, here we complement and further develop the results presented in~\cite{byrne1996importance} by systematically assessing the pattern formation potential of different stress-strain constitutive equations for the ECM within a mechanical model of pattern formation in biological tissues~\cite{byrne1996importance,murray1988mechanochemical,oster1983mechanical}. Compared to the work of~\cite{byrne1996importance} here we consider a wider range of constitutive models, we allow cell traction forces to be reduced by cell-cell contact inhibition, and undertake numerical simulations of the model equations showing the formation of cellular patterns both in one and in two spatial dimensions. A related study has been conducted by \cite{alonso2017mechanochemical}, who considered a mathematical model of pattern formation in the cell cytoplasm. 

The paper is structured as follows. In Section~\ref{sec:viscoelastic}, we recall the essentials of viscoelastic materials and provide a brief summary of the one-dimensional stress-strain constitutive equations that we examine. In Section~\ref{sec:model}, we describe the one-dimensional mechanical model of pattern formation in biological tissues that is used in our study, which follows closely the one considered in~\cite{byrne1996importance,murray1988mechanochemical,oster1983mechanical}. In Section~\ref{sec:lsa}, we carry out a linear stability analysis (LSA) of a biologically relevant homogeneous steady state of the model equations, derive dispersion relations when different stress-strain constitutive equations for the ECM are used, and investigate how the model parameters affect the dispersion relations obtained. In Section~\ref{sec:numerical}, we verify key results of LSA via numerical simulations of the model equations. In Section~\ref{sec:2d}, we complement these findings with the results of numerical simulations of a two-dimensional version of the mechanical model of pattern formation considered in the previous sections. Section~\ref{sec:discussion} concludes the paper and provides an overview of possible research perspectives.

\section{Essentials of viscoelastic materials and stress-strain constitutive equations}
\label{sec:viscoelastic}
In this section, we first recall the main properties of viscoelastic materials (see Section~\ref{sec:properties}). Then, we briefly present the one-dimensional stress-strain constitutive equations that are considered in our study and summarise the main rheological properties of linear viscoelastic materials that they capture (see Section~\ref{sec:constitutive}). Most of the contents of this section can be found in standard textbooks, such as~\cite{findley2013creep} [chapters 1 and 5] and \cite{mase1970continuum}, and are reported here for the sake of completeness. Specific considerations of and applications to living tissues can be found in \cite{fung1993}. 

\subsection{Essentials of viscoelastic materials}\label{sec:properties}
As the name suggests, viscoelastic materials exhibit both viscous and elastic characteristics, and the interplay between them may result in a wide range of rheological properties that can be examined through creep and stress relaxation tests. During a creep test, a constant stress is first applied to a specimen of material and then removed, and the time dynamic of the correspondent strain is tracked. During a stress relaxation test, a constant strain is imposed on a specimen of material and the evolution in time of the induced stress is observed~\cite{findley2013creep}.

Here we list the main properties of viscoelastic materials that may be observed during the first phase of a creep test (see properties 1a-1c), during the recovery phase, that is, when the constant stress is removed from the specimen (see properties 2a-2c), and during a stress relaxation test (see property 3).
 \\
\begin{enumerate}[label=(\alph*)]
\item[1a] {\it Instantaneous elasticity.} As soon as a stress is applied, an instantaneous corresponding strain is observed. 
\item[1b] {\it Delayed elasticity.} While the instantaneous elastic response to a stress is a purely elastic behaviour, due to the viscous nature of the material a delayed elastic response may also be observed. In this case, under constant stress the strain slowly and continuously increases at decreasing rate. 
\item[1c] {\it Viscous flow.} In some viscoelastic materials, under a constant stress, the strain continues to grow within the viscoelastic regime (\textit{i.e.} before plastic deformation). In particular, viscous flow occurs when the strain increases linearly with time and stops growing at removal of the stress only.
\item[2a] {\it Instantaneous recovery.} When the stress is removed, an instantaneous recovery (\textit{i.e.} an instantaneous strain decrease) is observed because of the elastic nature of the material.
\item[2b] {\it Delayed recovery.} Upon removal of the stress, a delayed recovery (\textit{i.e.} a continuous decrease of the strain at decreasing rate) occurs.
\item[2c] {\it Permanent set.} While elastic strain is reversible, in viscoelastic materials a non-zero strain{, known as ``permanent set'' or ``residual strain'',} may persist even when the stress is removed. 
\item[3\phantom{a}] {\it Stress relaxation.} Under constant strain, gradual relaxation of the induced stress occurs. In some cases, this may even culminate in total stress relaxation (\textit{i.e.} the stress decays to zero). 
\end{enumerate}
\bigskip
\noindent {The subset of these properties exhibited by a viscoelastic material will depend on  -- 
and hence define -- the type of material being tested. Moreover, during each phase of the creep test, more than one of the above properties may be observed. For instance, a Maxwell material under constant stress will exhibit instantaneous elasticity followed by viscous flow. }

\subsection{One-dimensional stress-strain constitutive equations examined in our study}\label{sec:constitutive}
\begin{figure}
\centering
\includegraphics[width=0.8\linewidth]{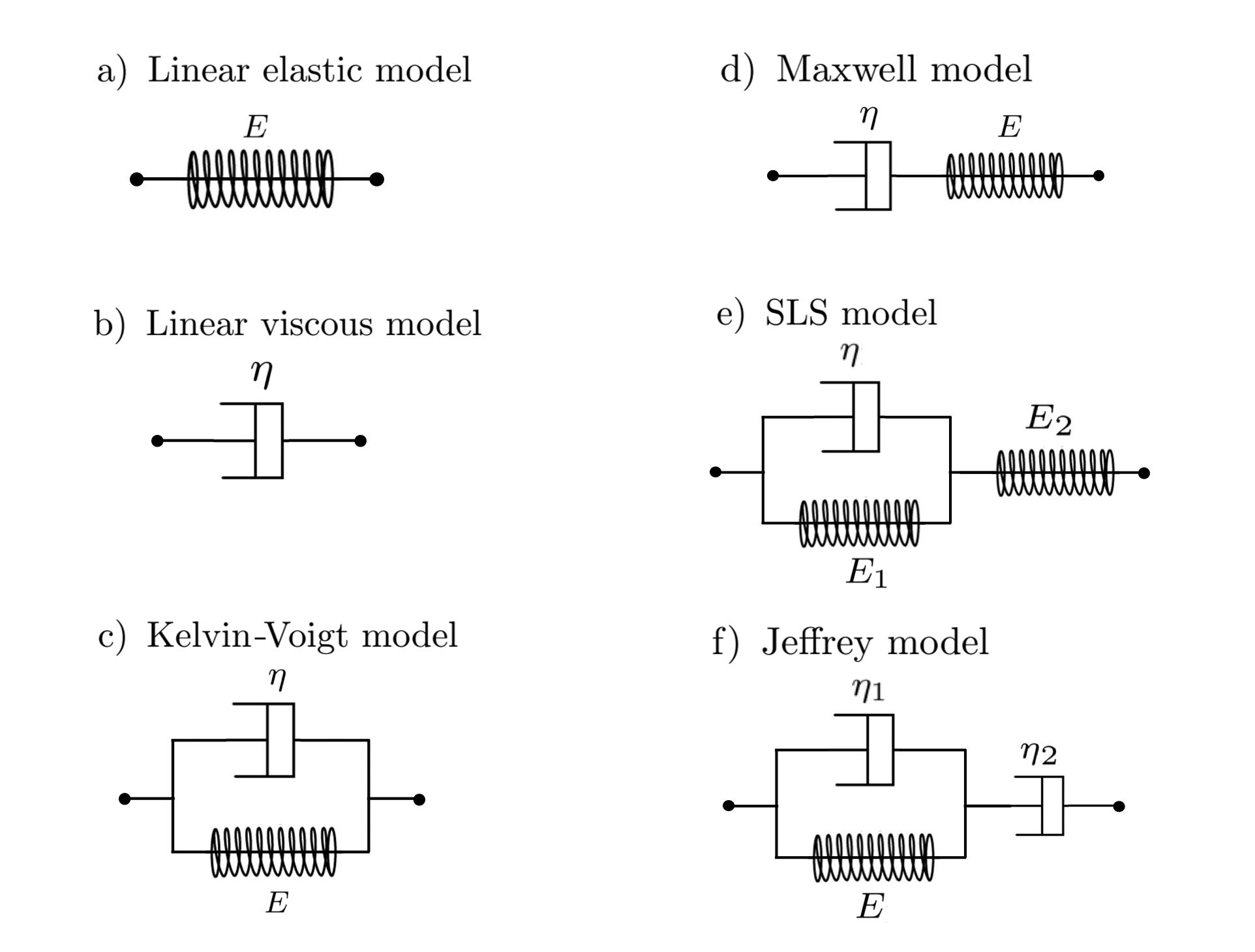}
\caption{\label{VEmodels}{Combinations of elastic springs and viscous dampers, together with the associated elastic ($E$, $E_1$, $E_2$) and viscous moduli ($\eta$, $\eta_1$, $\eta_2$), for the models of linear viscoelasticity considered in this work: the linear elastic model (a), the linear viscous model (b), the Kelvin-Voigt model (c), the Maxwell model (d), the SLS model (e), and the Jeffrey model (f).}}
\end{figure}

In this section, we briefly describe the different constitutive equations that are used in our study to represent the stress-strain relation of the ECM. In general, these equations can be used to predict how a viscoelastic material will react to different loading conditions, in one spatial dimension, and rely on the assumption that viscous and elastic characteristics of the material can be modelled, respectively, via linear combinations of dashpots and springs{, as illustrated in Figure~\ref{VEmodels}}. Different stress-strain constitutive equations correspond to different arrangements of these elements and capture different subsets of the rheological properties summarised in the previous section (see Table~\ref{tab:properties}). In the remainder of this section, we will denote the stress and the strain at position $x$ and time $t$ by $\s(t,x)$ and $\e(t,x)$, respectively.  

\paragraph{Linear elastic model.} When viscous characteristics are neglected, a linear viscoelastic material can be modelled as a purely elastic spring with elastic modulus (\emph{i.e.} Young's modulus) $E>0${, as illustrated in Figure~\ref{VEmodels}a}. In this case, the stress-strain constitutive equation is given by Hooke's spring law for continuous media, that is,
\begin{equation}
\s = E \e \,.
\label{ce:s}
\end{equation}

\paragraph{Linear viscous model.} When elastic characteristics are neglected, a linear viscoelastic material can be modelled as a purely viscous damper of viscosity $\eta >0${, as illustrated in Figure~\ref{VEmodels}b}. In this case, the stress-strain constitutive equation is given by Newton's law of viscosity, that is,
\begin{equation}
\s = \eta \, \ed \,.
\label{ce:d}
\end{equation}

\paragraph{Kelvin-Voigt model.} The Kelvin-Voigt model, {also known as the Voigt model,} relies on the assumption that viscous and elastic characteristics of a linear viscoelastic material can simultaneously be captured by considering a purely elastic spring with elastic modulus $E$ and a purely viscous damper of viscosity $\eta$ in parallel{, as illustrated in Figure~\ref{VEmodels}c}. The corresponding stress-strain constitutive equation is
\begin{equation}
\s = E \e + \eta \, \ed \,.
\label{ce:kv}
\end{equation}

\paragraph{Maxwell model.} The Maxwell model relies on the assumption that viscous and elastic characteristics of a linear viscoelastic material can be captured by considering a purely elastic spring with elastic modulus $E$ and a purely viscous damper of viscosity $\eta$ in series{, as illustrated in Figure~\ref{VEmodels}d}. The corresponding stress-strain constitutive equation is
\begin{equation}
\frac{1}{E} \,\sd+ \frac{\s}{\eta} = \ed \,.
\label{ce:m}
\end{equation}

\paragraph{Standard linear solid (SLS) model.} The SLS model, {also known as the Kelvin model,} relies on the assumption that viscous and elastic characteristics of a linear viscoelastic material can be captured by considering a Kelvin arm of elastic modulus $E_1$ and viscosity $\eta$ in series with a purely elastic spring of elastic modulus $E_2${, as illustrated in Figure~\ref{VEmodels}e}. 
The corresponding stress-strain constitutive equation is~\cite{mase1970continuum}
\begin{equation}
\frac{1}{E_2}\,\sd + \frac{1}{\eta} \bigg( 1+\frac{E_1}{E_2} \bigg) \s = \ed + \frac{E_1}{\eta}\e \,.
\label{ce:sls}
\end{equation}

\paragraph{{Jeffrey} model.} {The {Jeffrey} model, also known as the Oldroyd-B or 3-parameter viscous model,} relies on the assumption that viscous and elastic characteristics of a linear viscoelastic material can be captured by considering a Kelvin arm of elastic modulus $E$ and viscosity $\eta_1$ in series with a purely viscous damper of viscosity $\eta_2${, as illustrated in Figure~\ref{VEmodels}f}. 
The corresponding stress-strain constitutive equation is
\begin{equation}
\bigg( 1+\frac{\eta_1}{\eta_2} \bigg) \sd + \frac{E}{\eta_2}\s = \eta_1 \edd + E \ed \, .
\label{ce:3p}
\end{equation}

\paragraph{Generic 4-parameter model.} The following stress-strain constitutive equation encompasses all constitutive models of linear viscoelasticity whereby a combination of purely elastic springs and purely viscous dampers, up to a total of four elements, and therefore 4 parameters, is considered
\begin{equation}
a_2\sdd + a_1\sd + a_0\s = b_2\edd + b_1\ed + b_0\e \, .
\label{ce:4p}
\end{equation}
Here the non-negative, real parameters $a_0, a_1, a_2, b_0, b_1, b_2$ depend on the elastic moduli and the viscosities of the underlying combinations of springs and dampers. When these parameters are defined as in Table~\ref{tab:parameters}, the generic 4-parameter constitutive model~\eqref{ce:4p} reduces to the specific stress-strain constitutive equations~\eqref{ce:s}-\eqref{ce:3p}. For convenience of notation, we define the differential operators
\begin{equation}
\La_a := a_2 \dtt + a_1 \dt + a_0 \quad \text{and} \quad \La_b := b_2 \dtt + b_1 \dt + b_0
\label{La}
\end{equation}
so that the stress-strain constitutive equation~\eqref{ce:4p} can be rewritten in the following compact form
\begin{equation}
\La_a[\,\s\,] = \La_b[\,\e\,] \,.
\label{ce:4p2}
\end{equation}

{\setlength{\extrarowheight}{7pt}
\begin{table}[]
\caption{Relations between the generic 4-parameter model~\eqref{ce:4p} and the stress-strain constitutive equations~\eqref{ce:s}-\eqref{ce:3p}.}\label{tab:parameters}
\centering
\begin{tabular}{|l|c|c|c|c|c|c|}
\hline
\bf{Generic 4-parameters model} & $\bf{a_2}$ & $\bf{a_1}$ & $\bf{a_0}$ & $\bf{b_2}$ & $\bf{b_1}$ & $\bf{b_0}$ \\ [5pt] \hline 
Linear elastic model & 0 & 0 & 1 & 0 & 0 & $E$   \\ [5pt] \hline
Linear viscous model & 0 & 0 & 1 & 0 & $\eta$ & 0   \\ [5pt] \hline
Kelvin-Voigt model & 0 & 0 & 1 & 0 & $\eta$ & $E$   \\ [5pt] \hline
Maxwell model & 0 & $\frac{1}{E}$ & $\frac{1}{\eta}$ & 0 & 1 & 0   \\ [5pt] \hline
SLS model    & 0 & $\frac{1}{E_2}$ & $\frac{1}{\eta} \Big( 1+\frac{E_1}{E_2} \Big)$ & 0 & 1 & $\frac{E_1}{\eta}$   \\ [5pt] \hline
{Jeffrey} model & 0 & $1+\frac{\eta_1}{\eta_2}$ & $\frac{E}{\eta_2}$ & $\eta_1$ & $E$ & 0   \\[5pt]  \hline
\end{tabular}
\end{table}
}

A summary of the rheological properties of linear viscoelastic materials listed in Section~\ref{sec:properties} that are captured by the one-dimensional stress-strain constitutive equations~\eqref{ce:s}-\eqref{ce:3p} is provided in Table~\ref{tab:properties}. These properties can be examined through mathematical procedures that mimic creep and stress relaxation tests~\cite{findley2013creep}. Notice that, for all these constitutive models, instantaneous elasticity correlates with instantaneous recovery, delayed elasticity correlates with delayed recovery, and viscous flow correlates with permanent set. 
{Materials are said to be more solid-like when their elastic response dominates their viscous response, and more fluid-like in the opposite case~\cite{nargess2021form}. For this reason, models of linear viscoelasticity that capture viscous flow and, as a consequence, permanent set -- such as the Maxwell model and the Jeffrey model -- are classified as ``fluid-like models'', while those which do not -- such as the Kelvin-Voigt model and the SLS model -- are classified as ``solid-like models''. In the remainder of the paper we are going to include the linear viscous model in the fluid-like class and the linear elastic model in the solid-like class, as they capture -- or do not capture -- the relevant properties, even if they are not models of viscoelasticity \textit{per se}.}

{\setlength{\extrarowheight}{7pt}
\begin{table}[]
\caption{Properties of linear viscoelastic materials captured by the stress-strain constitutive equations~\eqref{ce:s}-\eqref{ce:3p}.}\label{tab:properties}
\centering
\begin{tabular}{|l|c|c|c|c|c|c|c|} %
\hline
 & \parbox[c]{2cm}{ Instantaneous  \\ elasticity} & \parbox[c]{1.3cm}{Delayed\\elasticity} & \parbox[c]{1cm}{Viscous \\flow} & \parbox[c]{2cm}{Instantaneous \\ recovery} & \parbox[c]{1.1cm}{Delayed \\ recovery} & \parbox[c]{1.6cm}{Permanent \\ set} & \parbox[c]{1.4cm}{Stress \\ relaxation} \\ [5pt] \hline 
Linear elastic model & \checkmark &  &  & \checkmark &  &  & \    \\ [5pt] \hline
Linear viscous model & &  & \checkmark &  &  & \checkmark & N. A.     \\ [5pt] \hline
Kelvin-Voigt model &  & \checkmark & & & \checkmark &  &    \\ [5pt] \hline
Maxwell model & \checkmark &  & \checkmark & \checkmark &  & \checkmark & \checkmark    \\ [5pt] \hline
SLS model    & \checkmark & \checkmark &  & \checkmark & \checkmark &  & \checkmark   \\ [5pt] \hline
{Jeffrey} model &  & \checkmark & \checkmark & & \checkmark & \checkmark & \checkmark  \\[5pt]  \hline
\end{tabular}
\end{table}
} 

\section{A one-dimensional mechanical model of pattern formation} \label{sec:model}
We consider a one-dimensional region of tissue and represent the normalised densities of cells and ECM at time $t\in[0,T]$ and position $x \in [\ell,L]$ by means of the non-negative functions $n(t,x)$ and $\rho(t,x)$, respectively. We let $u(t,x)$ model the displacement of a material point of the cell-ECM system originally at position $x$, which is induced by mechanical interactions between cells and the ECM -- \emph{i.e.} cells pull on the ECM in which they are embedded, thus inducing ECM compression and densification which in turn cause a passive form of cell repositioning~\cite{van2018mechanoreciprocity}.

\subsection{Dynamics of the cells}
Following~\cite{murray1988mechanochemical,oster1983mechanical}, we consider a scenario where cells change their position according to a combination of: (i) undirected, random movement, which we describe through Fick's first law of diffusion with diffusivity (\emph{i.e.} cell motility) $D>0$; (ii) haptotaxis (\emph{i.e.} cell movement up the density gradient of the ECM) with haptotactic sensitivity $\alpha>0$; (iii) passive repositioning caused by mechanical interactions between cells and the ECM, which is modelled as an advection with velocity field $\dt u$. Moreover, we model variation of the normalised cell density caused by cell proliferation and death via logistic growth with intrinsic growth rate $r>0$ and unitary local carrying capacity. Under these assumptions, we describe cell dynamics through the following balance equation for $n(t,x)$
\begin{equation}
\dt n  \, = \, {  \partial_x \left[ \,D\,\partial_x n \,- n \, \left( \alpha \, \partial_x \rho + \dt u \right)\right] }+ r \, n(1-n)
\label{eq:n}
\end{equation}
subject to suitable initial and boundary conditions.

\subsection{Dynamics of the ECM}
As was done for the cell dynamics, in a similar manner we model compression and densification of the ECM induced by cell-ECM interactions as an advection with velocity field $\dt u$. Furthermore, as in~\cite{murray1988mechanochemical,oster1983mechanical}, we neglect secretion of ECM components by the cells since this process occurs on a slower time scale compared to mechanical interactions between cells and the ECM. Under these assumptions, we describe the cell dynamics through the following transport equation for $\rho(t,x)$  
\begin{equation}
\dt \rho  \, = \,  - \partial_x \left(\rho \, \dt u \right)
\label{eq:rho}
\end{equation}
subject to suitable initial and boundary conditions.

\subsection{Force-balance equation for the cell-ECM system}
Following~\cite{murray1988mechanochemical,oster1983mechanical}, we represent the cell-ECM system as a linear viscoelastic material with low Reynolds number (\emph{i.e.} inertial terms are negligible compared to viscous terms) and we assume the cell-ECM system to be in mechanical equilibrium (\emph{i.e.} traction forces generated by the cells are in mechanical equilibrium with viscoelastic restoring forces developed in the ECM and any other external forces). Under these assumptions, the force-balance equation for the cell-ECM system is of the form
\begin{equation}
\partial_x \left(\sigma_c + \sigma_m\right) +\rho \, F = 0 \, ,
\label{eq:fbeg}
\end{equation}
where $\sigma_m(t,x)$ is the contribution to the stress of the cell-ECM system coming from the ECM, $\sigma_c(t,x)$ is the contribution to the stress of the cell-ECM system coming from the cells, and $F(t,x)$ is the external force per unit matrix density, which comes from the surrounding tissue that constitutes the underlying substratum to which the ECM is attached. 

The stress $\sigma_c$ is related to cellular traction forces acting on the ECM and is defined as
\begin{equation}
\s_c := \tau \, f(n) \, n \, {\big(\, \rho + \beta\, \partial^2_{xx}\rho\, \big) } \quad \text{with} \quad f(n) := \dfrac{1}{1+\lambda \, n^2} \,.
\label{eq:sc}
\end{equation}
Definition~\eqref{eq:sc} relies on the assumption that the stress generated by cell traction on the ECM is proportional to the cell density $n$ and {-- in the short range --} the ECM density $\rho${, while the term $\beta\, \partial^2_{xx}\rho$ accounts for long-range cell traction effects, with $\beta$ being the long-range traction proportionality constant.} The factor of proportionality is given by a positive parameter, $\tau$, which measures the average traction force generated by a cell, multiplied by a non-negative and monotonically decreasing function of the cell density, $f(n)$, which models the fact that the average traction force generated by a cell is reduced by cell-cell contact inhibition~\cite{murray2001mathematical}. The parameter $\lambda \geq 0$ measures the level of cell traction force inhibition and assuming $\lambda=0$ corresponds to neglecting the reduction in the cell traction forces caused by cellular crowding. 

The stress $\sigma_m$ is given by the stress-strain constitutive equation that is used for the ECM, which we choose to be the general constitutive model~\eqref{ce:4p2} with the strain $\e(t,x)$ being given by the gradient of the displacement $u(t,x)$, that is, $\e=\dx u$. Therefore, we define the stress-strain relation of the ECM via the following equation
\begin{equation}
\La_a[\,\sigma_m\,] = \La_b[\,\dx u\,] \,,
\label{eq:sm}
\end{equation}
where the differential operators $\La_a$ and $\La_b$ are defined according to~\eqref{La}.

Assuming the surrounding tissue to which the ECM is attached to be a linear elastic material~\cite{murray2001mathematical}, the external body force $F$ can be modelled as a restoring force proportional to the cell-ECM displacement, that is, 
\begin{equation}
F := - s \, u \,  .
\label{eq:def}
\end{equation}
Here the parameter $s>0$ represents the elastic modulus of the surrounding tissue.

In order to obtain a closed equation for the displacement $u(t,x)$, we apply the differential operator $\La_a[\,\cdot\,]$ to the force-balance equation~\eqref{eq:fbeg} and then substitute~\eqref{eq:sc}-\eqref{eq:def} into the resulting equation. In so doing, we find 
\begin{align*}
 &\La_a\,[\, \dx\, (\s_m +\s_c) \,] = - \La_a\,[\, \rho \, F\,] \\
&{\Leftrightarrow}\, \La_a\,[\, \dx\, \s_m \,] \,+\, \La_a\,[\, \dx\, \s_c  \,] \,= \, \La_a\,[\, s\rho u\,]  \\
&{\Leftrightarrow}\, \dx \, \La_a\,[\,  \s_m \,] \,= \,  \La_a\,[\, s\rho u\,] \,-\, \La_a\,[\, \dx\, \s_c  \,] \\
&{\Leftrightarrow}\, \dx \, \La_b\,[\, \dx u \,] \, =  \,  \La_a\,[\,  s\rho u \,-\, \dx \s_c \,] \\
&{\Leftrightarrow}\, \La_b\,[\,  \dxx u  \,] \,=  \,  \La_a\,[\,  s\rho u \,-\, \dx \s_c \,]  \,,\\
\end{align*}
that is,
\begin{equation}
\La_b\,[\,  \dxx u  \,] \,=  \,  \La_a\,\left[\,  s\rho u \,-\, \dx \left( \frac{\tau n}{1+\lambda n^2} {(\rho + \beta \dxx \rho)} \right) \,\right]  \,.
\label{eq:fb}
\end{equation}
Finally, to close the system, equation~\eqref{eq:fb} needs to be supplied with suitable initial and boundary conditions.

\subsection{Boundary conditions}
We close our mechanical model of pattern formation defined by the system of PDEs~\eqref{eq:n}, \eqref{eq:rho} and~\eqref{eq:fb} with the following boundary conditions 
\begin{equation}
\label{bc_1d}
\begin{cases}
&n(t,\ell) = n(t,L) \,, \quad \dx n(t,\ell) = \dx n(t,L) \,, \\[5pt] 
&\rho(t,\ell) = \rho(t,L) \,, {\quad \dxx \rho(t,\ell) = \dxx \rho(t,L)\,, } \\[5pt] 
&u(t,\ell) = u(t,L) \,, \quad \dx u(t,\ell) = \dx u(t,L) \,,\\[10pt] 
\end{cases}
\qquad \text{for all } t\in {[}0,T] \,.
\end{equation}
{Here, the conditions on the derivatives of $n$, $\rho$ and $u$ ensure that the fluxes in equations~\eqref{eq:n} and~\eqref{eq:rho}, and the overall stress $(\s_m +\s_c)$ in equation~\eqref{eq:fb}, are periodic on the boundary, \textit{i.e.} they ensure that
\begin{equation*}
\begin{cases}
&\left[ \,D\,\partial_x n \,- n \, \left( \alpha \, \partial_x \rho + \dt u \right)\right]_{x=\ell} \,=\, \left[ \,D\,\partial_x n \,- n \, \left( \alpha \, \partial_x \rho + \dt u \right)\right]_{x=L} \,, \\[6pt]
&\left[ n \,  \dt u \right]_{x=\ell} \,=\, \left[ n \,  \dt u \right]_{x=L} \,, \\[5pt]
&\Big[ \tau \dfrac{ n}{(1+\lambda^2)} \, (\, \rho + \beta\, \partial^2_{xx}\rho\, )  + \s_m \Big]_{x=\ell} \,=\, \Big[ \tau  \dfrac{ n}{(1+\lambda^2)} \, (\, \rho + \beta\, \partial^2_{xx}\rho\, )  + \s_m \Big]_{x=L} \,,
\end{cases} 
\quad \text{for all } t\in [0,T] \,,
\end{equation*}
with $\s_m$ given as a function of $\dx u$ in equation~\eqref{eq:sm}, according to the selected constitutive model.} The periodic boundary conditions~\eqref{bc_1d} reproduce a biological scenario in which the spatial region considered is part of a larger area of tissue whereby similar dynamics of the cells and the ECM occur.

\section{Linear stability analysis and dispersion relations}\label{sec:lsa}
In this section, we carry out LSA of a biologically relevant homogeneous steady state of the system of PDEs~\eqref{eq:n}, \eqref{eq:rho} and~\eqref{eq:fb} (see Section~\ref{sec:LSA}) and we compare the dispersion relations obtained when the constitutive models~\eqref{ce:s}-\eqref{ce:3p} are alternatively used to represent the contribution to the overall stress coming from the ECM, in order to explore the pattern formation potential of these stress-strain constitutive equations (see Section~\ref{disprelplots}).

\subsection{Linear stability analysis}
\label{sec:LSA}
\paragraph{Biologically relevant homogeneous steady state.} 
All non-trivial homogeneous steady states $(\bar{n},\bar{\rho},\bar{u})^\intercal$ of the system of PDEs~\eqref{eq:n}, \eqref{eq:rho} and~\eqref{eq:fb} subject to boundary conditions~\eqref{bc_1d} have components $\bn \equiv 1$ and $\bu \equiv  0$, and { we consider the arbitrary non-trivial steady state $\br \equiv \rho_0> 0$} amongst the infinite number of possible homogeneous steady states of the transport equation~\eqref{eq:rho} for the normalised ECM density $\rho$.
Hence, we focus our attention on the biologically relevant homogeneous steady state ${\bf{\bar{v}}}=(1,{\rho_0},0)^\intercal$.

\paragraph{Linear stability analysis to spatially homogeneous perturbations.} In order to undertake linear stability analysis of the steady state ${\bf{\bar{v}}}=(1,{\rho_0},0)^\intercal$ to spatially homogeneous perturbations, we make the ansatz ${\bf{v}}(t,x) \equiv {\bf{\bar{v}}} + {\bf{\tilde{v}}}(t)$, where the vector ${\bf{\tilde{v}}} (t)= (\tilde{n}(t),\tilde{\rho}(t),\tilde{u}(t))^\intercal$ models small spatially homogeneous perturbations, and linearise the system of PDEs~\eqref{eq:n}, \eqref{eq:rho} and~\eqref{eq:fb} about the steady state ${\bf{\bar{v}}}$. Assuming $\tilde{n}(t)$, $\tilde{\rho}(t)$ and $\tilde{u}(t)$ to be proportional to $\exp{(\psi t)}$, with $\psi \neq 0$, one can  easily verify that $\psi$ satisfies the algebraic equation $\psi(\psi +r)(\psi^2 a_2 +\psi a_1 +a_0)=0$. Since $r$ is positive and the parameters $a_0$, $a_1$ and $a_2$ are all non-negative, the solution $\psi$ of such an algebraic equation is necessarily negative and, therefore, the small perturbations $\tilde{n}(t)$, $\tilde{\rho}(t)$ and $\tilde{u}(t)$ will decay to zero as $t \to \infty$. This implies that the steady state ${\bf{\bar{v}}}$ will be stable to spatially homogeneous perturbations for any choice of the parameter $a_0$, $a_1$, $a_2$, $b_0$, $b_1$ and $b_2$ in the stress-strain constitutive equation~\eqref{eq:sm} (\emph{i.e.} for all constitutive models~\eqref{ce:s}-\eqref{ce:3p}).

\paragraph{Linear stability analysis to spatially inhomogeneous perturbations.}  In order to undertake linear stability analysis of the steady state ${\bf{\bar{v}}}=(1,{\rho_0},0)^\intercal$ to spatially inhomogeneous perturbations, we make the ansatz ${\bf{v}}(t,x) = {\bf{\bar{v}}} + {\bf{\tilde{v}}}(t,x)$, where the vector ${\bf{\tilde{v}}} (t,x)= (\tilde{n}(t,x),\tilde{\rho}(t,x),\tilde{u}(t,x))^\intercal$ models small spatially inhomogeneous perturbations, and linearise the system of PDEs~\eqref{eq:n}, \eqref{eq:rho} and~\eqref{eq:fb} about the steady state ${\bf{\bar{v}}}$. Assuming $\tilde{n}(t,x)$, $\tilde{\rho}(t,x)$ and $\tilde{u}(t,x)$ to be proportional to $\exp{(\psi t + i k x)}$, with $\psi \neq 0$ and $k \neq 0$, we find that $\psi$ satisfies the following equation
\begin{equation}
 \label{full}
{\psi \, \Big[\,} c_3(k^2)\psi^3 + c_2(k^2)\psi^2 +c_1(k^2)\psi +c_0(k^2) \,{\Big]} = 0 \, ,
\end{equation}
with
\begin{equation}
\label{c3}
c_3(k^2) := {a_2\tau\lambda_1\beta\, k^4\,} +\, \big[b_2 -a_2\tau(\lambda_1+\lambda_2{\rho_0})\big]\,k^2 \,+\, a_2s{\rho_0}
\end{equation}
\begin{equation}
\label{c2}
\begin{aligned}
c_2(k^2) := &\; {a_2 \tau\lambda_1D\beta\, k^6} \, + \, \big[ b_2D - {a_2\tau(\lambda_2\rho_0\alpha+D\lambda_1-r\lambda_1\beta) +a_1\tau\lambda_1\beta} \big] \,k^4 \, \\
& +\, \big[  b_2r + b_1 + a_2(Ds{\rho_0-r\tau\lambda_1})-a_1\tau(\lambda_1+\lambda_2{\rho_0})\big] \,k^2\, + (a_1+a_2r)s{\rho_0}
\end{aligned}
\end{equation}
\begin{equation}
\label{c1}
\begin{aligned}
c_1(k^2) := &\; {a_1 \tau\lambda_1D\beta\, k^6} \, + \, \big[ b_1D - {a_1\tau(\lambda_2\rho_0\alpha+D\lambda_1-r\lambda_1\beta) +a_0\tau\lambda_1\beta \big]} \,k^4 \, \\
& +\, \big[  b_1r + b_0 +a_1(Ds{\rho_0-r\tau\lambda_1})-a_0\tau(\lambda_1+\lambda_2{\rho_0})\big] \,k^2\, + (a_0+a_1r)s{\rho_0}
\end{aligned}
\end{equation}
and
\begin{equation}
\label{c0}
\begin{aligned}
c_0(k^2) := &\; {a_0 \tau\lambda_1D\beta\, k^6} \, + \, \big[ b_0D - {a_0\tau(\lambda_2\rho_0\alpha+D\lambda_1-r\lambda_1\beta) }\big] \,k^4 \, \\
& +\, \big[  b_0r +a_0(Ds{\rho_0-r\tau\lambda_1})\big] \,k^2\, + a_0rs{\rho_0}
\end{aligned}
\end{equation}
where
$$
\lambda_1 := \dfrac{1}{1+\lambda}  \quad \text{and} \quad \lambda_2 := \dfrac{(1-\lambda)}{(1+\lambda)^2} \ .
$$

{Equation~\eqref{full} has multiple solutions $(\psi(k^2))$ for each $k^2$ and we denote by ${\rm Re}(\cdot)$ the maximum real part of all these solutions.}
For cell patterns to emerge, we need the non-trivial homogeneous steady state ${\bf{\bar{v}}}$ to {be unstable to} spatially inhomogeneous perturbations, that is, we need ${\rm Re}(\psi(k^2))>0$ for some $k^2>0$. Notice that a necessary condition for this to happen is that at least one amongst $c_0(k^2)$, $c_1(k^2)$, $c_2(k^2)$ and $c_3(k^2)$ is negative for some $k^2>0$. Hence, the fact that if $\tau=0$ then $c_0(k^2)$, $c_1(k^2)$, $c_2(k^2)$ and $c_3(k^2)$ are all non-negative for any value of $k^2$ allows us to conclude that having $\tau>0$ is a necessary condition for pattern formation to occur.  This was expected based on the results presented in~\cite{murray2001mathematical} and references therein.

In the case where the model parameters are such that $c_2(k^2) =0$ and $c_3(k^2) = 0$, solving equation~\eqref{full} for $\psi$ gives the following dispersion relation
\begin{equation} \label{lin}
\psi(k^2) = -\frac{c_0(k^2)}{c_1(k^2)} 
\end{equation}
and for the condition ${\rm Re}(\psi(k^2))>0$ to be met it suffices that{, for some $k^2>0$,}
$$
c_0(k^2)>0 \quad \text{and} \quad c_1(k^2)<0 \quad \text{or} \quad c_0(k^2)<0 \quad \text{and} \quad c_1(k^2)>0 \, .
$$

On the other hand, when the model parameters are such that only $c_3(k^2) = 0$, from equation~\eqref{full} we obtain the following dispersion relation
\begin{equation} \label{quad}
\psi(k^2) = \frac{-c_1(k^2) \pm \sqrt{\big(c_1(k^2)\big)^2 - 4c_2(k^2)c_0(k^2)}}{2c_2(k^2)} \,,
\end{equation}
and for the condition ${\rm Re}(\psi(k^2))>0$ to be satisfied it is sufficient that one of the following four sets of conditions holds
$$
c_2(k^2)>0 \quad \text{and} \quad c_0(k^2)<0  \quad \text{or} \quad c_2(k^2)>0 \,, \quad c_1(k^2)<0 \quad \text{and} \quad c_0(k^2)>0
$$
or
$$
c_2(k^2)<0 \quad \text{and} \quad c_0(k^2)>0 \quad \text{or} \quad c_2(k^2)<0 \,, \quad c_1(k^2)>0 \quad \text{and} \quad c_0(k^2)<0 \ .
$$

Finally, in the general case where the model parameters are such that $c_3(k^2) \neq 0$ as well, from equation~\eqref{full} we obtain the following dispersion relation
\begin{equation} \label{cub}
\psi(k^2) = \left\{ q + \left[ q^2 + \left( m - p^2 \right)^3 \right]^{1/2} \right\}^{1/3} + \, \left\{ q - \left[ q^2 + \left( m - p^2 \right)^3 \right]^{1/2} \right\}^{1/3}  + \, p \,,
\end{equation}
where $p \equiv p(k^2)$, $q \equiv q(k^2)$ and $m \equiv m(k^2)$ are defined as
$$
p := -\frac{c_2}{3c_3} \,, \quad q := p^3 + \frac{c_2c_1 - 3c_3c_0}{6c_3^2} \,, \quad m := \frac{c_1}{3c_3} \,.
$$
In this case, identifying sufficient conditions to ensure that the real part of $\psi(k^2)$ is positive for some $k^2>0$ requires lengthy algebraic calculations. We refer the interested reader to~\cite{gilmore2012mechanochemical}, where the Routh-Hurwitz stability criterion was used to analyse this general case and obtain more explicit conditions on the model parameters under which pattern formation occurs.

\subsection{Dispersion relations}\label{disprelplots}
Substituting the definitions of $a_0$, $a_1$, $a_2$, $b_0$, $b_1$ and $b_2$ corresponding to the stress-strain constitutive equations~\eqref{ce:s}-\eqref{ce:3p}, which are reported in Table~\ref{tab:parameters}, into definitions~\eqref{c3}-\eqref{c0} for $c_0(k^2)$, $c_1(k^2)$, $c_2(k^2)$ and $c_3(k^2)$, and then using the dispersion relation {given by formula}~\eqref{lin}, \eqref{quad} or \eqref{cub} depending on the values of $c_2(k^2)$ and $c_3(k^2)$ so obtained, we derive the dispersion relation for each of the constitutive models~\eqref{ce:s}-\eqref{ce:3p}. {In particular, we are interested in whether the real part of each dispersion relation is positive, so whenever multiple roots are calculated -- for instance using~\eqref{quad} -- the largest root is considered. In addition, dispersion relations throughout this section are plotted against the quantity $k/\pi$, which directly correlates with perturbation modes and can therefore better highlight mode selection during the sensitivity analysis. }

\begin{figure}[htb!]
 \centering
  \includegraphics[width=0.9\linewidth]{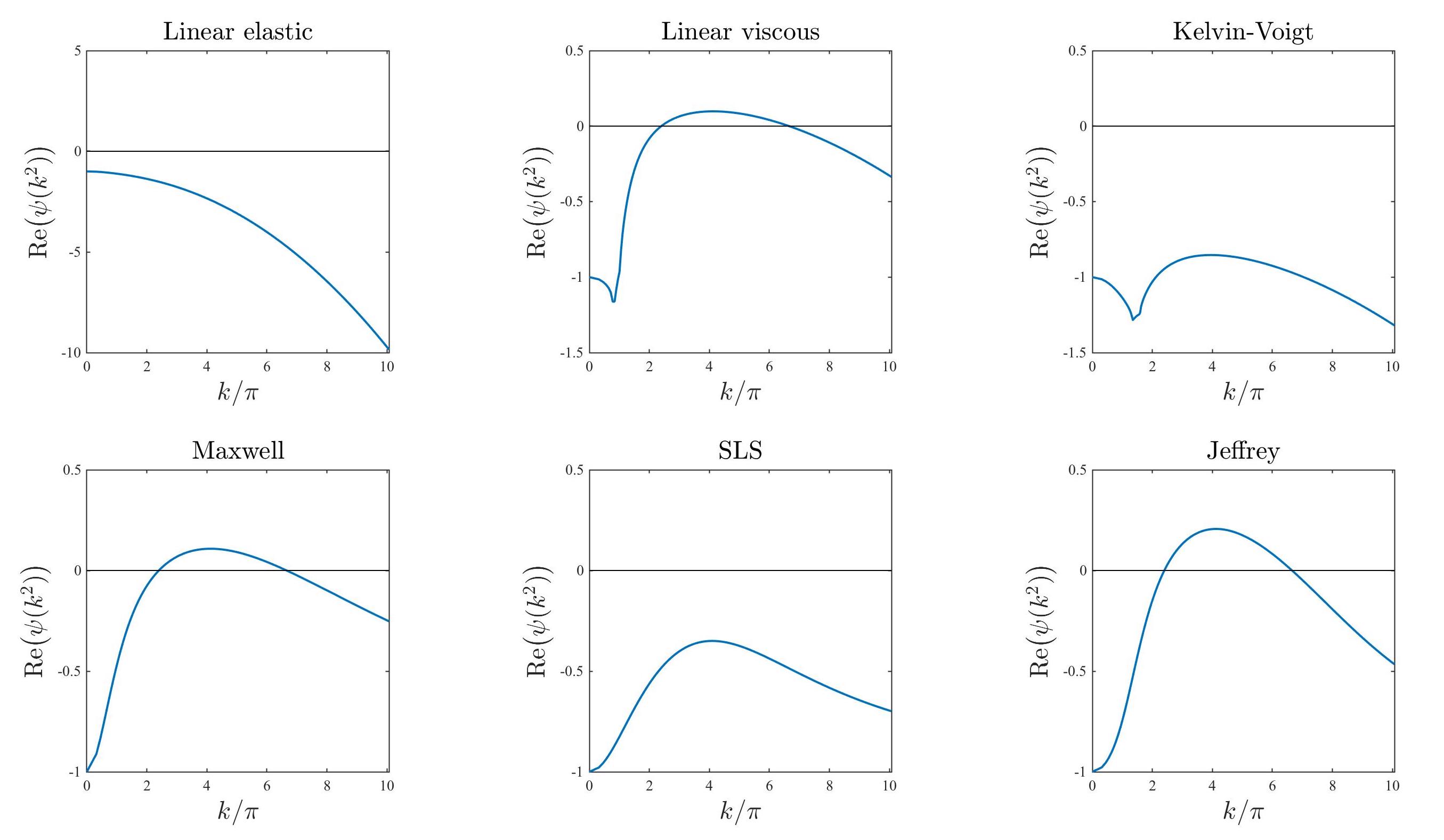}
\caption{{\bf Base-case dispersion relations.} Dispersion relations corresponding to the stress-strain constitutive equations~\eqref{ce:s}-\eqref{ce:3p} for the base-case set of parameter values given by~\eqref{eq:param1} and~\eqref{eq:param2}.}
\label{DispRel0}
\end{figure}

\paragraph{Base-case dispersion relations.} Figure~\ref{DispRel0} displays the dispersion relations obtained for the stress-strain constitutive equations~\eqref{ce:s}-\eqref{ce:3p} under the following base-case parameter values
\begin{equation}
\label{eq:param1}
{ E=1\,, \quad E_1=E_2=\frac{1}{2}E=0.5 \,, \quad \eta=1\,, \quad  \eta_1=\eta_2=\frac{1}{2}\eta =0.5 \,, \quad D=0.01 \,, }
\end{equation}
\begin{equation}
\label{eq:param2}
{\rho_0=1 \,, \quad \alpha=0.05 \,, \quad r=1 \,, \quad s=10 \,, \quad \lambda=0.5 \,, \quad \tau=0.2 \, \quad \beta=0.005 \,.}
\end{equation}
The parameter values given by~\eqref{eq:param1} and~\eqref{eq:param2} are chosen for illustrative purposes, in order to highlight the different qualitative behaviour of the dispersion relations obtained using different models{, and are comparable with nondimensional parameter values that can be found in the extant literature (see Appendix~\ref{appendix:par} for further details).}
 A comparison between the plots in Figure~\ref{DispRel0} reveals that {fluid-like models}, that is, the linear viscous model~\eqref{ce:d}, the Maxwell model~\eqref{ce:m} and the {Jeffrey} model~\eqref{ce:3p} (\emph{cf.} Table~\ref{tab:properties}), have a higher pattern formation potential than {solid-like models}, {since under the same parameter set they exhibit a range -- or, more precisely, they exhibit the same range -- of unstable modes (\textit{i.e.} ${\rm Re}(\psi(k^2))>0$ for a range of values of $k/\pi$), while the others have no unstable modes.}

We now undertake a sensitivity analysis with respect to the different model parameters and discuss key changes that occur in the base-case dispersion relations displayed in Figure~\ref{DispRel0}.

\begin{figure}[htb!]
 \centering
  \includegraphics[width=.72\linewidth]{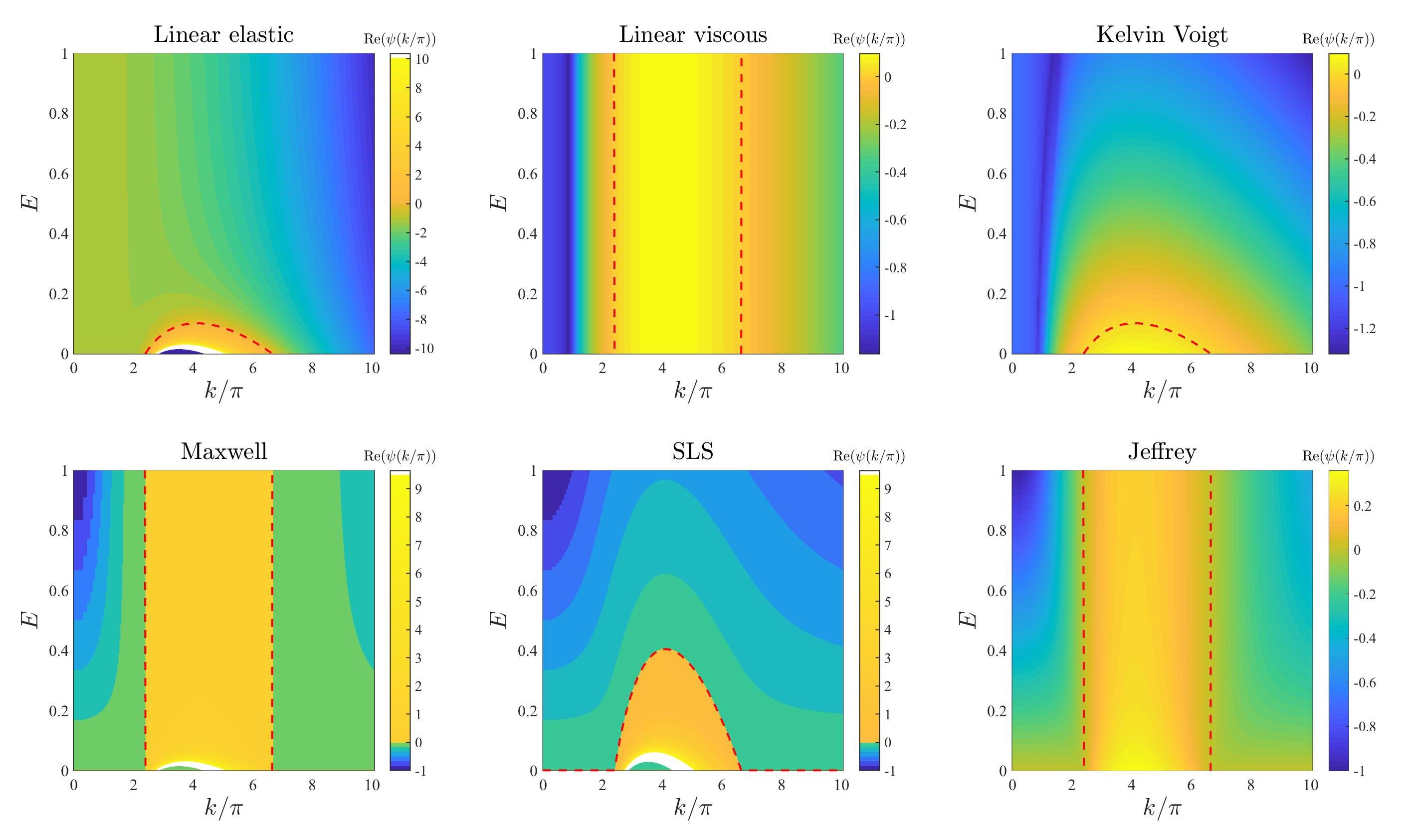}
\caption{{\bf Effects of varying the ECM elasticity.}  Dispersion relations corresponding to the stress-strain constitutive equations~\eqref{ce:s}-\eqref{ce:3p} for increasing values of the ECM elasticity, that is {for $E\in[0,1]$}. The values of the other parameters are given by~\eqref{eq:param1} and~\eqref{eq:param2}. {White regions in the plots related to the linear elastic model, the Maxwell model and the SLS model correspond to ${\rm Re}(\psi(k^2))>10$ (\emph{i.e.} a vertical asymptote is present in the dispersion relation). Red dashed lines mark contour lines where ${\rm Re}(\psi(k^2))=0$.}}
\label{DispRele}
\end{figure}

\paragraph{ECM elasticity.} The plots in Figure~\ref{DispRele} illustrate how the base-case dispersion relations displayed in Figure~\ref{DispRel0} change when different values of the {parameter $E$, and therefore also $E_1$ and $E_2$} (\emph{i.e.} the parameters modelling ECM elasticity), are considered. 
{These plots show that lower values of these parameters correlate with overall larger values of ${\rm Re}(\psi(k^2))$ for all constitutive models, except for the linear viscous one, which corresponds to speeding up the formation of spatial patterns, when these may form. In addition, sufficiently small values of the parameters $E$, $E_1$ and $E_2$ allow the linear elastic model~\eqref{ce:s}, the Kelvin-Voigt model~\eqref{ce:kv}, and the SLS model~\eqref{ce:sls} to exhibit unstable modes. However, further lowering the values of these parameters} appears to lead to singular dispersion relations ({\it cf.} the plots for the linear elastic model~\eqref{ce:s}, the Maxwell model~\eqref{ce:m} and the SLS model~\eqref{ce:sls} in Figure~\ref{DispRele}), which suggests that linear stability theory may fail in the regime of low ECM elasticity.

\begin{center}
\begin{figure}[h!]
 \centering
  \includegraphics[width=.72\linewidth]{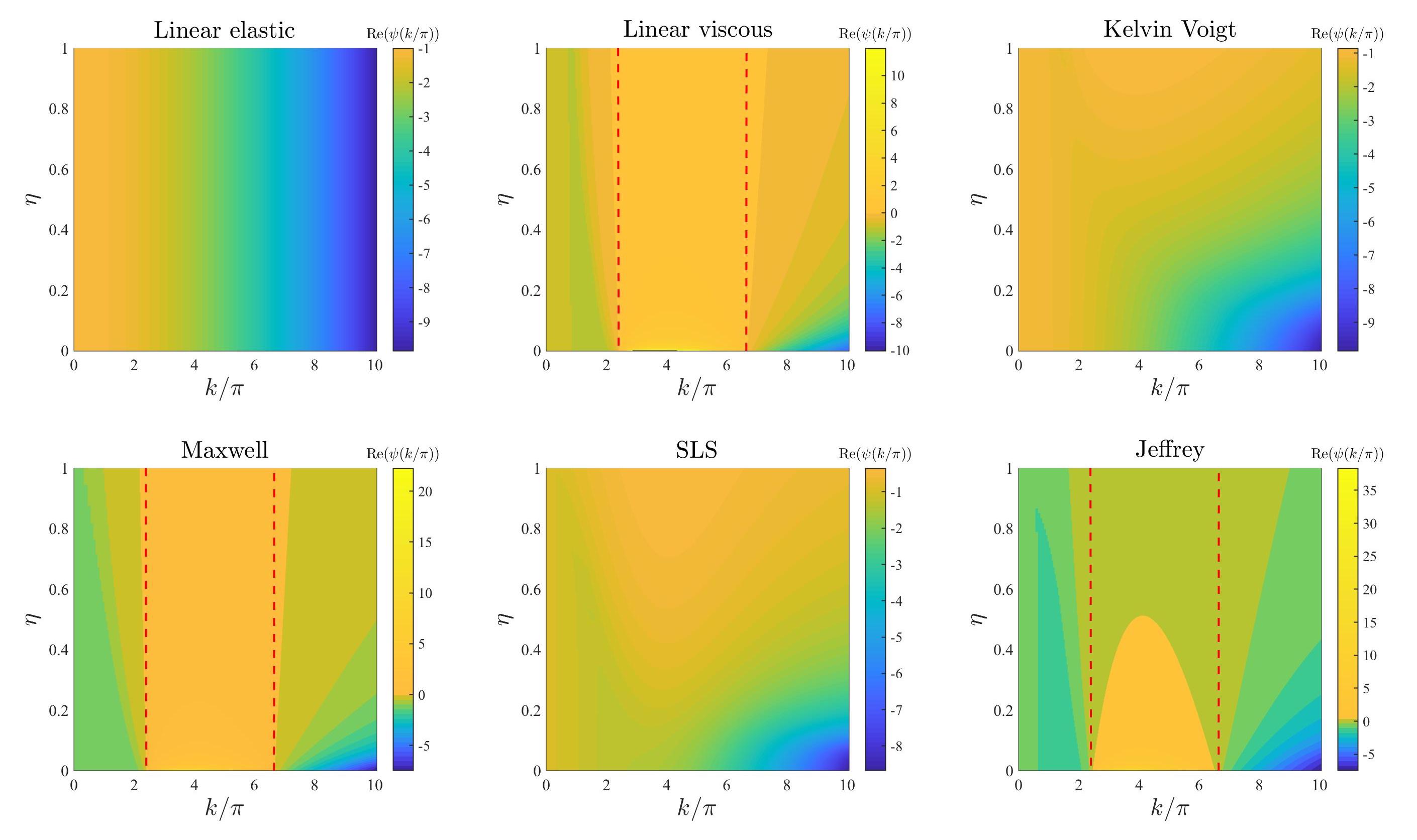}
\caption{{\bf Effects of varying the ECM viscosity.}  Dispersion relations corresponding to the stress-strain constitutive equations~\eqref{ce:s}-\eqref{ce:3p} for increasing values of the ECM viscosity, that is {for $\eta\in[0,1]$}. The values of the other parameters are given by~\eqref{eq:param1} and~\eqref{eq:param2}. Red dashed lines mark contour lines where ${\rm Re}(\psi(k^2))=0$.}
\label{DispReleta}
\end{figure}
\end{center}

\paragraph{ECM viscosity.} The plots in Figure~\ref{DispReleta} illustrate how the base-case dispersion relations displayed in Figure~\ref{DispRel0} change when different values of the {parameter $\eta$,  and therefore also $\eta_1$ and $\eta_2$} (\emph{i.e.} the parameters modelling ECM viscosity), are considered. These plots show that larger values of these parameters leave the range of {modes} for which ${\rm Re}(\psi(k^2))>0$ unchanged but reduce the values of ${\rm Re}(\psi(k^2))$. This supports the idea that a higher ECM viscosity may not change the pattern formation potential of the different constitutive models but may slow down the corresponding pattern formation processes.

\begin{figure}[h!]
 \centering
  \includegraphics[width=.72\linewidth]{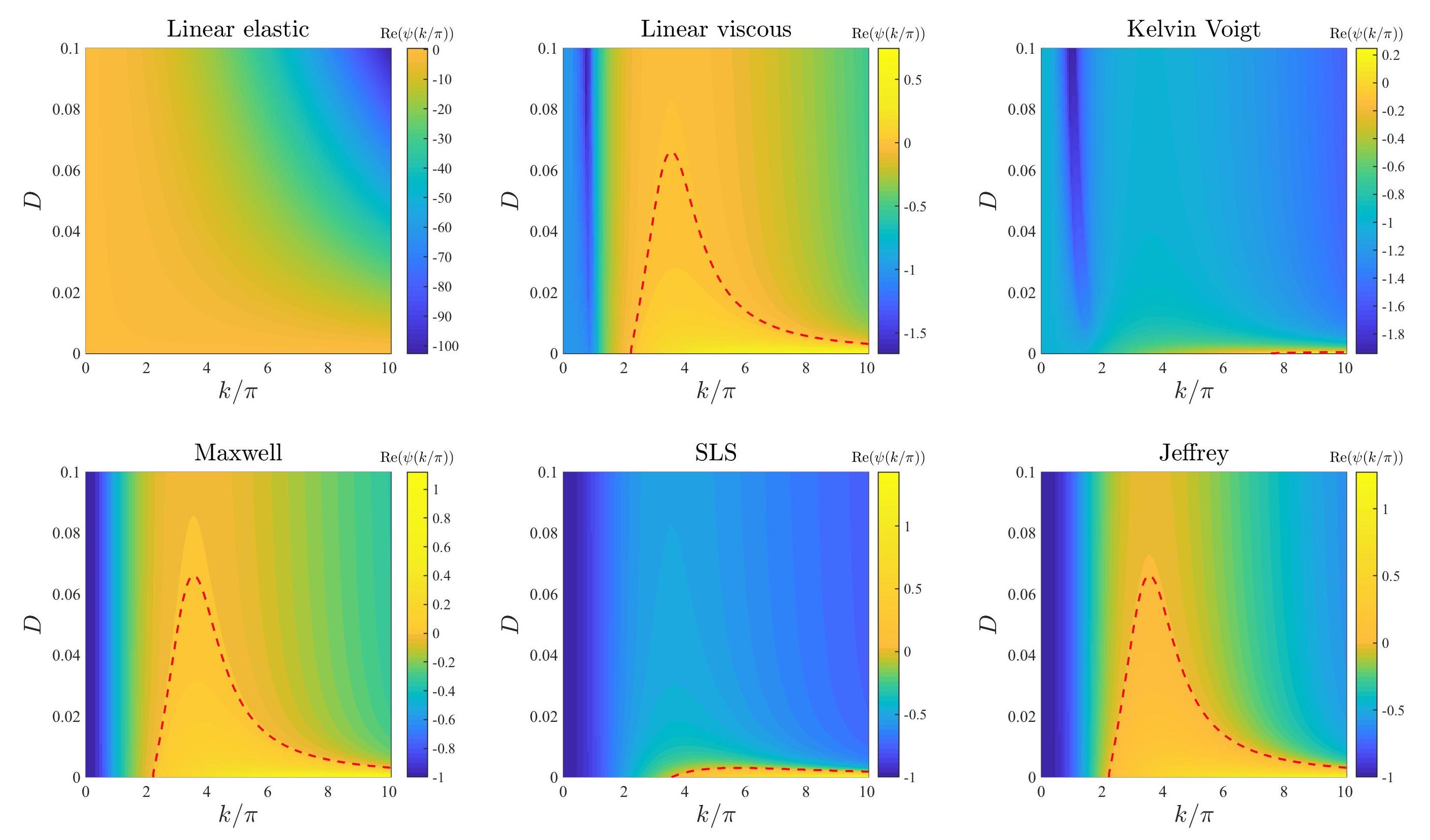}
\caption{{\bf Effects of varying the cell motility.} Dispersion relations corresponding to the stress-strain constitutive equations~\eqref{ce:s}-\eqref{ce:3p} for increasing values of the cell motility, that is {for $D\in[0,0.1]$}. The values of the other parameters are given by~\eqref{eq:param1} and~\eqref{eq:param2}. Red dashed lines mark contour lines where ${\rm Re}(\psi(k^2))=0$.}
\label{DispRelD}
\end{figure}

\paragraph{Cell motility.} The plots in Figure~\ref{DispRelD} illustrate how the base-case dispersion relations displayed in Figure~\ref{DispRel0} change when different values of the parameter $D$ (\emph{i.e.} the parameter modelling cell motility) are considered. These plots show that larger values of this parameter may significantly shrink the range of {modes} for which ${\rm Re}(\psi(k^2))>0$. In particular, {with the exception of the linear elastic model, all constitutive models exhibit: infinitely many unstable modes when $D \to 0$; a finite number of unstable modes for intermediate values of $D$; no unstable modes for sufficiently large values of $D$.} This is to be expected due to the stabilising effect of undirected, random cell movement and indicates that higher cell motility may correspond to lower pattern formation potential. 

\begin{figure}[h!]
 \centering
  \includegraphics[width=.72\linewidth]{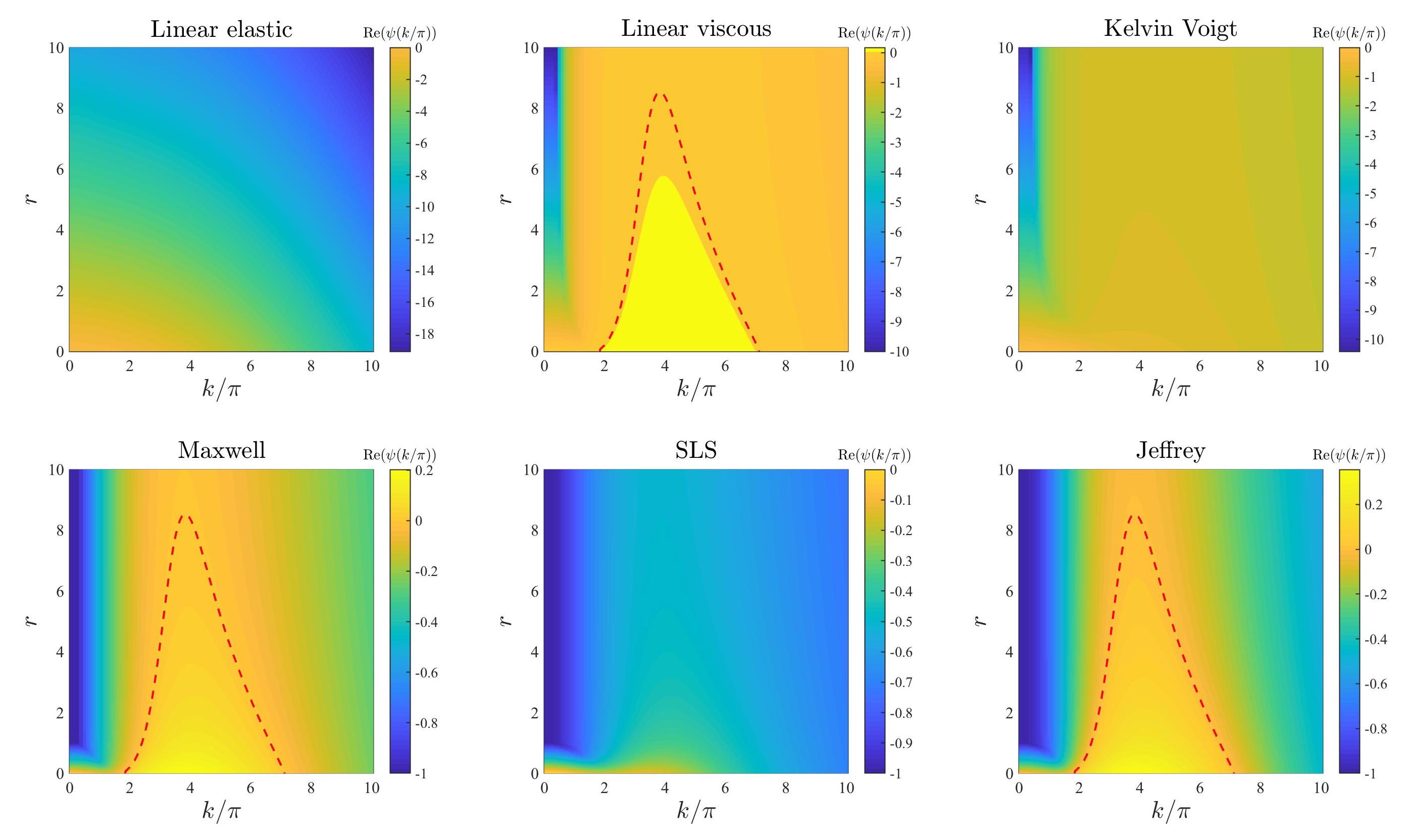}
\caption{{\bf Effects of varying the intrinsic growth rate of the cell density.} Dispersion relations corresponding to the stress-strain constitutive equations~\eqref{ce:s}-\eqref{ce:3p} for increasing values of the intrinsic growth rate of the cell density, that is {for $r\in[0,10]$}. The values of the other parameters are given by~\eqref{eq:param1} and~\eqref{eq:param2}. Red dashed lines mark contour lines where ${\rm Re}(\psi(k^2))=0$.}
\label{DispRelr}
\end{figure}

\begin{figure}[h!]
 \centering
  \includegraphics[width=.72\linewidth]{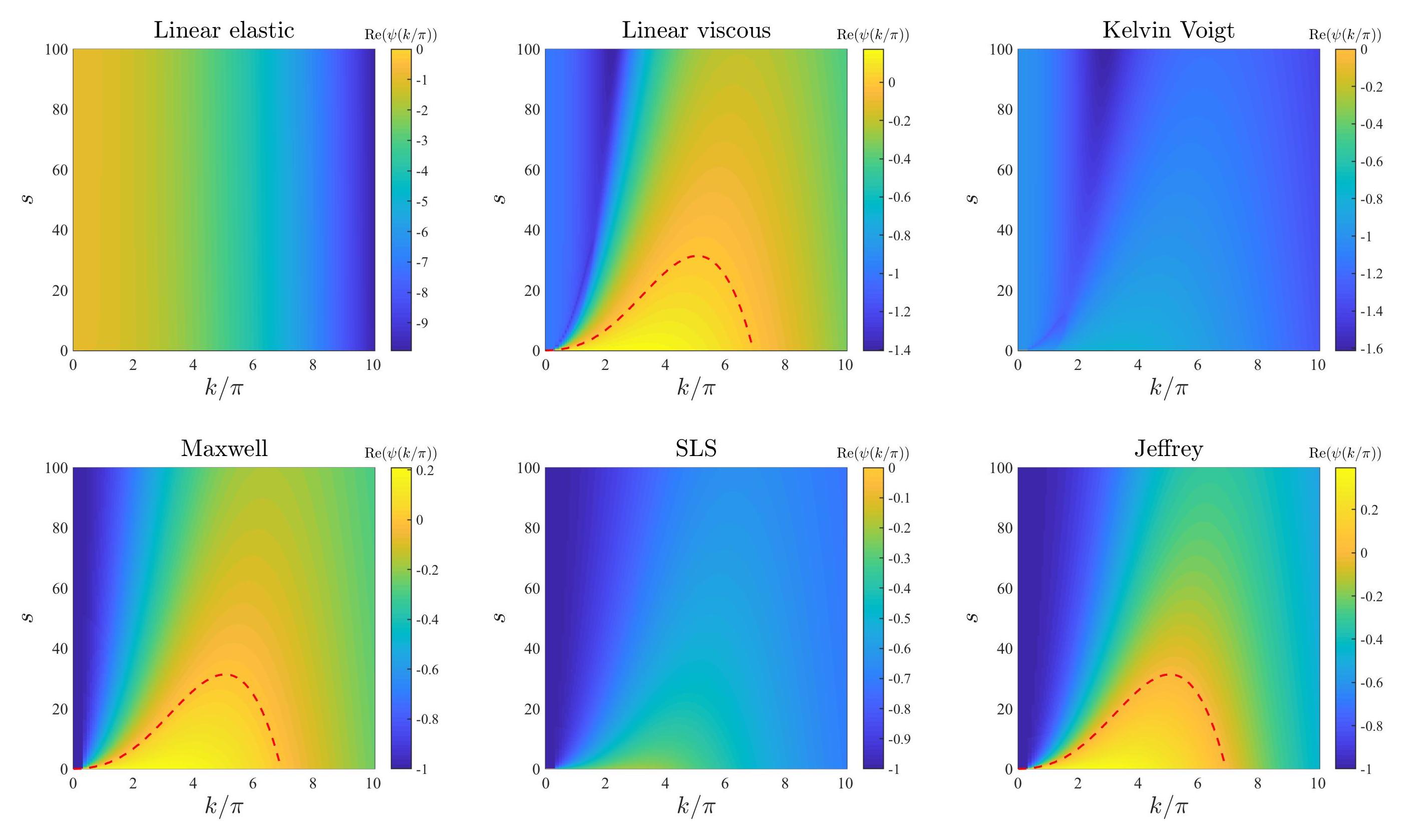}
\caption{{\bf Effects of varying the elasticity of the surrounding tissue.} Dispersion relations corresponding to the stress-strain constitutive equations~\eqref{ce:s}-\eqref{ce:3p} for increasing values of the elasticity of the surrounding tissue, that is {for $s\in[0,100]$}. The values of the other parameters are given by~\eqref{eq:param1} and~\eqref{eq:param2}. Red dashed lines mark contour lines where ${\rm Re}(\psi(k^2))=0$.}
\label{DispRels}
\end{figure}

\paragraph{Intrinsic growth rate of the cell density {and elasticity of the surrounding tissue}.} The plots in {Figures~\ref{DispRelr} and~\ref{DispRels}} illustrate how the base-case dispersion relations displayed in Figure~\ref{DispRel0} change when different values of the parameter $r$ (\emph{i.e.} the intrinsic growth rate of the cell density) {and the parameter $s$ (\emph{i.e.} the elasticity of the surrounding tissue)} are, respectively, considered. These plots show that considering larger values of {these parameters} reduces the values of ${\rm Re}(\psi(k^2))$ for {all constitutive models, and in particular} it shrinks the range of unstable modes for the {linear viscous model~\eqref{ce:d}, the Maxwell model~\eqref{ce:m} and the {Jeffrey} model~\eqref{ce:3p}, which can become stable for values of $r$ or $s$ sufficiently large.} This supports the idea that higher growth rates of the cell density (\emph{i.e.} faster cell proliferation and death){, and higher substrate elasticity (\emph{i.e.} stronger external tethering force)} may slow down pattern formation processes {and overall reduce the pattern formation potential for all constitutive models. Moreover, the plots in Figure~\ref{DispRels} indicate that higher values of $s$ may in particular reduce the pattern formation potential of the different constitutive models by making it more likely that ${\rm Re}(\psi(k^2))<0$ for smaller values of ${k/\pi}$ (\emph{i.e.} low-frequency perturbation modes will be more likely to vanish).}

\begin{figure}[h!]
 \centering
  \includegraphics[width=.72\linewidth]{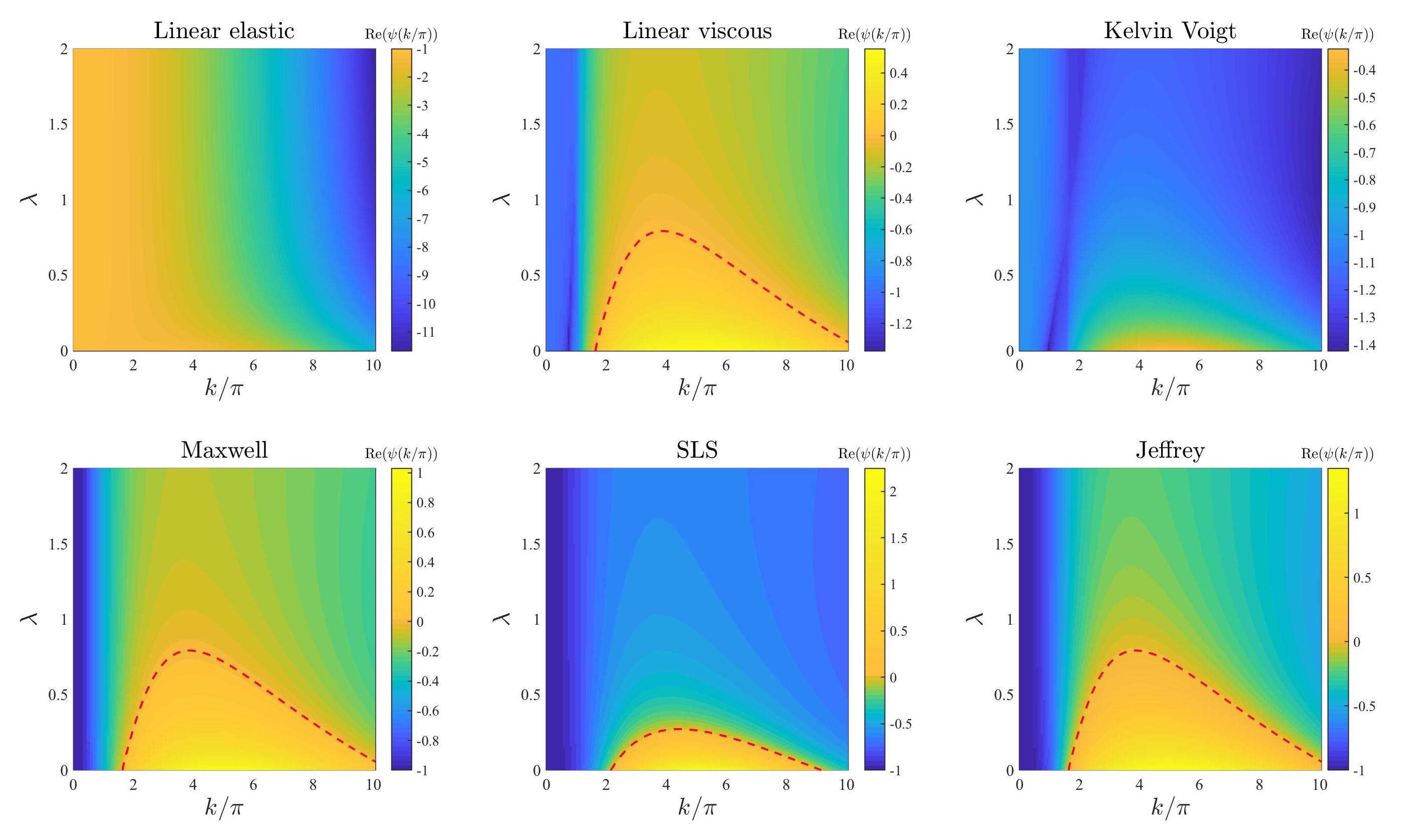}
\caption{{\bf Effects of varying the level of cell-cell contact inhibition of the cell traction forces.} Dispersion relations corresponding to the stress-strain constitutive equations~\eqref{ce:s}-\eqref{ce:3p} for increasing levels of cell-cell contact inhibition of the cell traction forces, that is {for $\lambda\in[0,2]$}. The values of the other parameters are given by~\eqref{eq:param1} and~\eqref{eq:param2}. Red dashed lines mark contour lines where ${\rm Re}(\psi(k^2))=0$.}
\label{DispRellambda}
\end{figure}

\begin{figure}[h!]
 \centering
  \includegraphics[width=.72\linewidth]{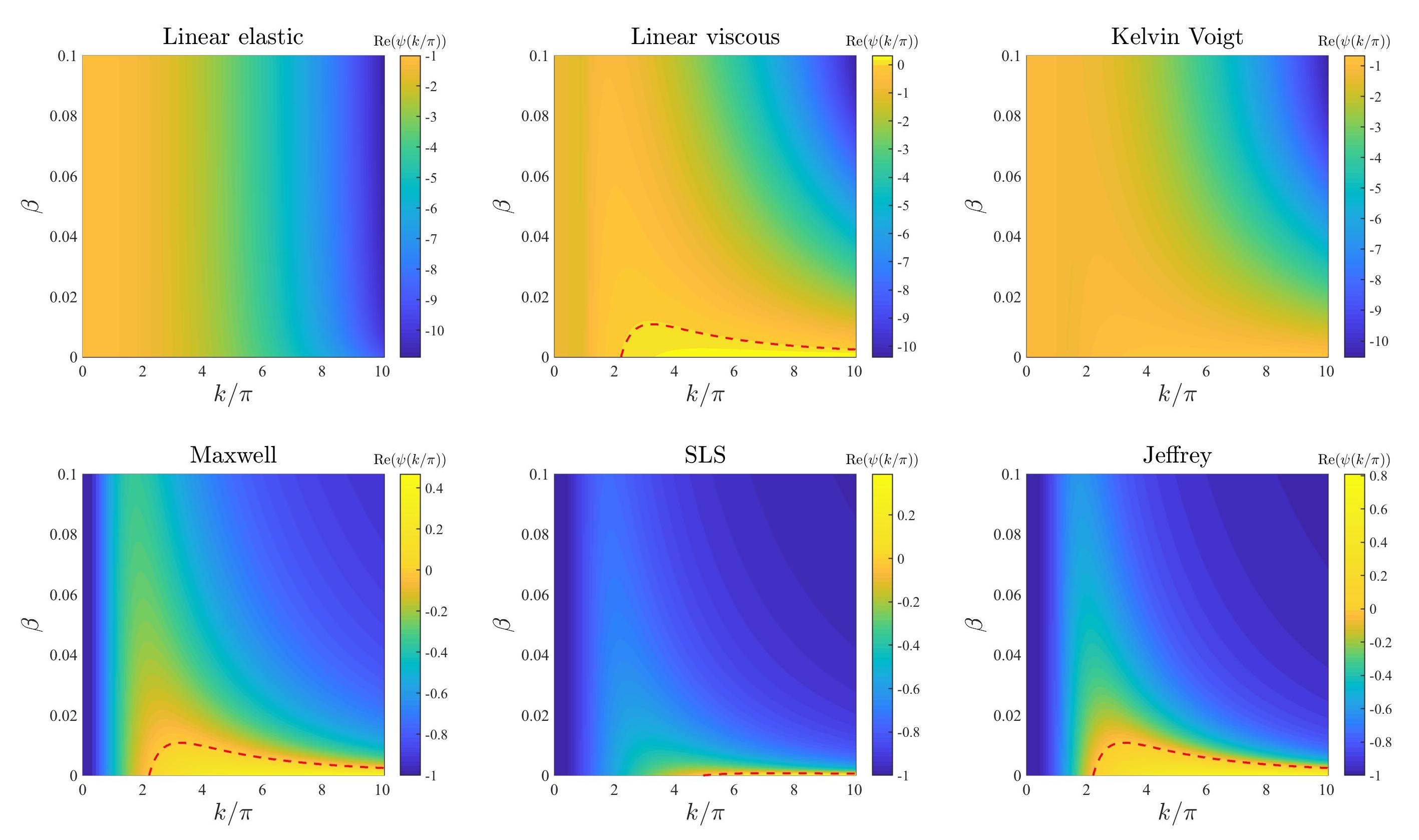}
\caption{{{\bf Effects of varying the long-range cell traction forces.} Dispersion relations corresponding to the stress-strain constitutive equations~\eqref{ce:s}-\eqref{ce:3p} for increasing long-range cell traction forces, that is for $\beta\in[0,0.1]$. The values of the other parameters are given by~\eqref{eq:param1} and~\eqref{eq:param2}. Red dashed lines mark contour lines where ${\rm Re}(\psi(k^2))=0$.}}
\label{DispRelbeta}
\end{figure}

{\paragraph{Level of contact inhibition of the cell traction forces and long-range cell traction forces.} The plots in Figures~\ref{DispRellambda} and~\ref{DispRelbeta} illustrate how the base-case dispersion relations displayed in Figure~\ref{DispRel0} change when different values of the parameter $\lambda$ (\emph{i.e.} the level of cell-cell contact inhibition of the cell traction forces) and the parameter $\beta$ (\emph{i.e.} the long-range cell traction forces) are, respectively, considered. Considerations similar to those previously made about the dispersion relations obtained for increasing values of the parameters $r$ and $s$ apply to the case where increasing values of the parameter $\lambda$ and the parameter $\beta$ are considered.  In addition to these considerations, the plots in Figures~\ref{DispRellambda} and~\ref{DispRelbeta} indicate that for small enough values of $\lambda$ or $\beta$ the SLS model~\eqref{ce:sls} can exhibit unstable modes, which further suggests that weaker contact inhibition of cell traction forces and lower long-range cell traction forces foster pattern formation. Moreover, the plots in Figure~\ref{DispRelbeta} indicate that in the asymptotic regime $\beta\to 0$ we may observe infinitely many unstable modes (\emph{i.e.} ${\rm Re}(\psi(k^2))>0$ for arbitrarily large wavenumebers), exiting the regime of physically meaningful pattern forming instabilities~\cite{moreo2010modelling,perelson1986nonlinear}. }

\begin{figure}[h!]
 \centering
  \includegraphics[width=.72\linewidth]{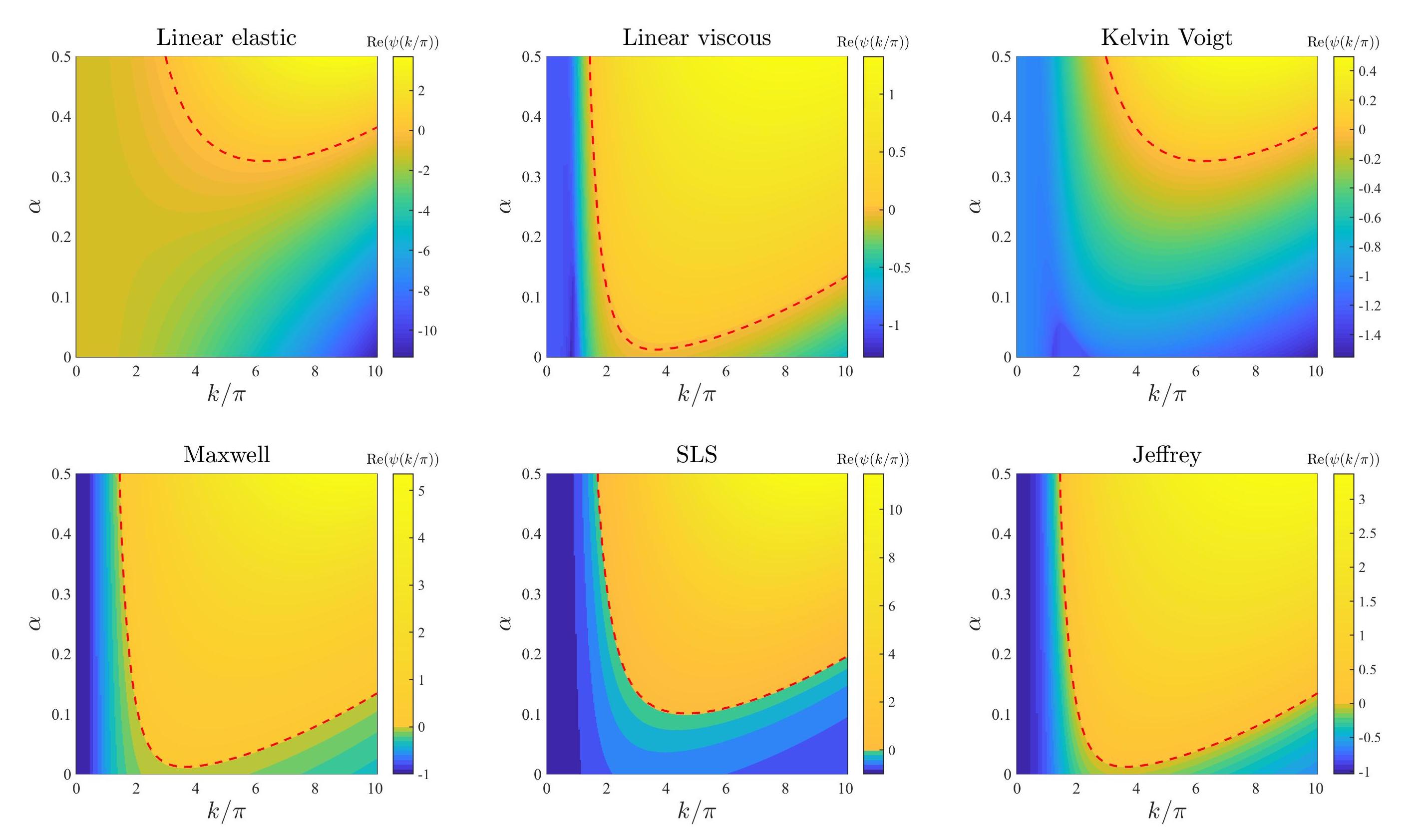}
\caption{{\bf Effects of varying the cell haptotactic sensitivity.} Dispersion relations corresponding to the stress-strain constitutive equations~\eqref{ce:s}-\eqref{ce:3p} for increasing values of the cell haptotactic sensitivity, that is {for $\alpha\in[0,0.5]$}. The values of the other parameters are given by~\eqref{eq:param1} and~\eqref{eq:param2}. Red dashed lines mark contour lines where ${\rm Re}(\psi(k^2))=0$.}
\label{DispRelalpha}
\end{figure}

\begin{figure}[h!]
 \centering
  \includegraphics[width=.72\linewidth]{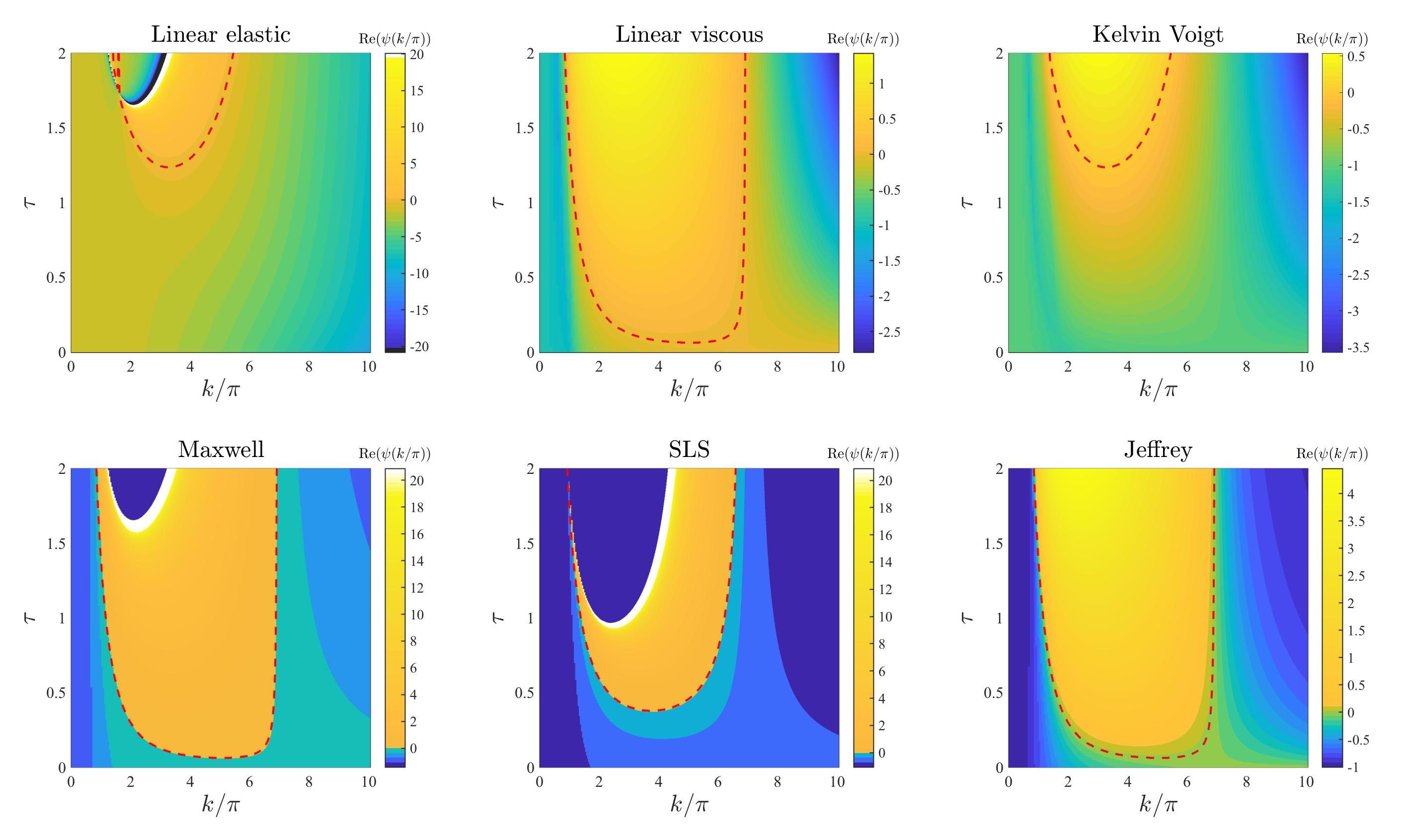}
\caption{{\bf Effects of varying the cell traction forces.} Dispersion relations corresponding to the stress-strain constitutive equations~\eqref{ce:s}-\eqref{ce:3p} for increasing cell traction forces, that is {for $\tau\in[0,2]$}. The values of the other parameters are given by~\eqref{eq:param1} and~\eqref{eq:param2}. {White and black regions in the plots related to the linear elastic model, the Maxwell model and the SLS model correspond, respectively, to ${\rm Re}(\psi(k^2))>20$ and ${\rm Re}(\psi(k^2))<-20$ (\emph{i.e.} a vertical asymptote is present in the dispersion relation). Red dashed lines mark contour lines where ${\rm Re}(\psi(k^2))=0$.}}
\label{DispReltau}
\end{figure}

\paragraph{Cell haptotactic sensitivity and cell traction forces.} The plots in Figures~\ref{DispRelalpha} and~\ref{DispReltau} illustrate how the base-case dispersion relations displayed in Figure~\ref{DispRel0} change when different values of the parameter $\alpha$ (\emph{i.e.} the cell haptotactic sensitivity) and the parameter $\tau$ (\emph{i.e.} the cell traction force) are, respectively, considered. As expected~\cite{murray2001mathematical}, larger values of these parameters {overall increase the value of ${\rm Re}(\psi(k^2))$ and broaden the range of values of modes for which ${\rm Re}(\psi(k^2))>0$, so that for large enough values of these parameters the linear viscous model~\eqref{ce:d}, the Kelvin-Voigt model~\eqref{ce:kv} and the SLS model~\eqref{ce:sls} can exhibit unstable modes. However, sufficiently large values of $\tau$ appear to lead to singular dispersion relations ({\it cf.} the plots for the linear elastic model~\eqref{ce:s}, the Maxwell model~\eqref{ce:m} and the SLS model~\eqref{ce:sls} in Figure~\ref{DispReltau}), which suggests that linear stability theory may fail in the regime of high cell traction for certain constitutive models, as previously observed in~\cite{byrne1996importance}.} 

\begin{figure}[h!]
 \centering
  \includegraphics[width=.72\linewidth]{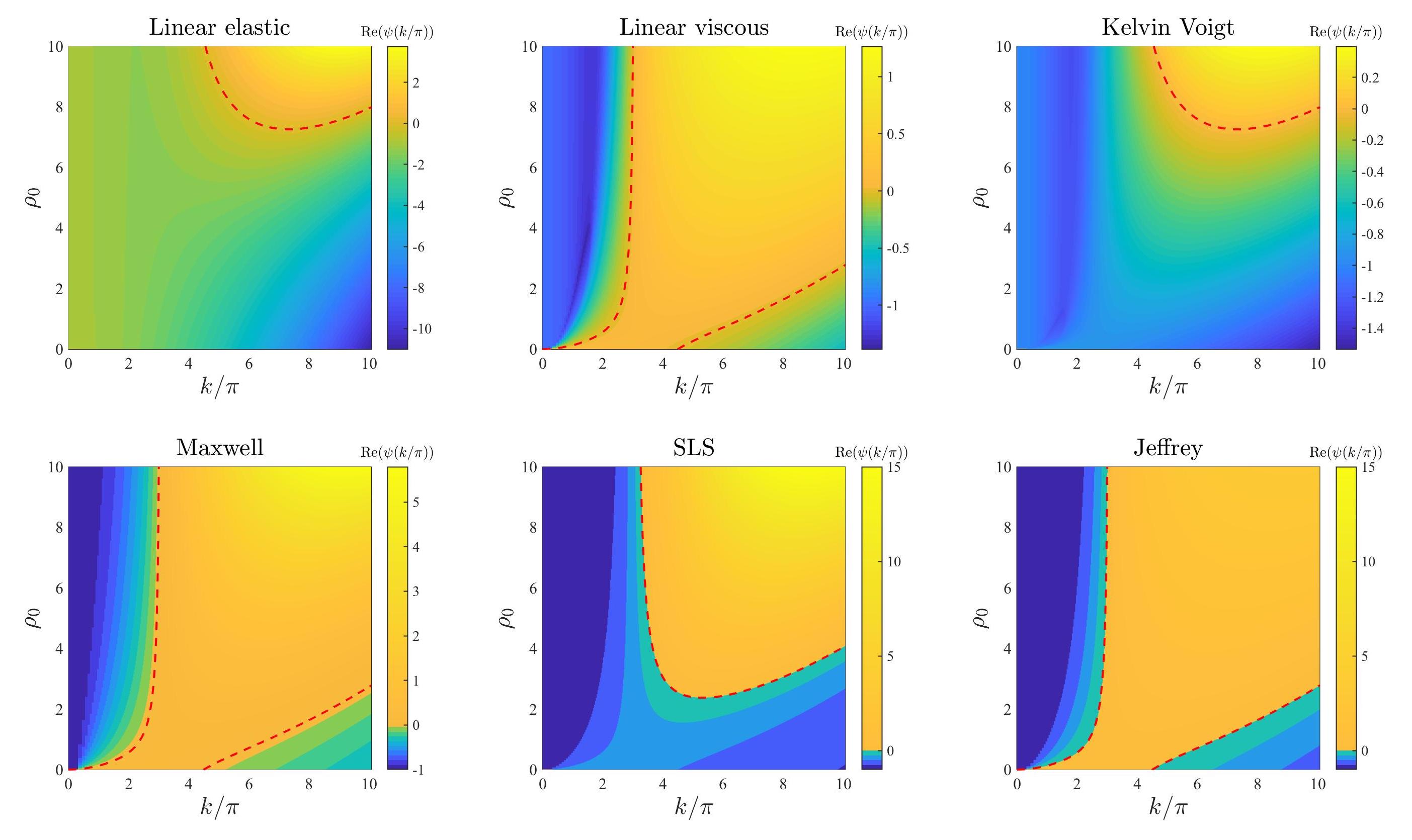}
\caption{{{\bf Effects of varying the initial ECM density.} Dispersion relations corresponding to the stress-strain constitutive equations~\eqref{ce:s}-\eqref{ce:3p} for increasing values of the initial ECM density, that is for $\rho_0\in[0,10]$. The values of the other parameters are given by~\eqref{eq:param1} and~\eqref{eq:param2}. Red dashed lines mark contour lines where ${\rm Re}(\psi(k^2))=0$.}}
\label{DispRelp0}
\end{figure}

{\paragraph{Initial ECM density.} The plots in Figure~\ref{DispRelp0} illustrate how the base-case dispersion relations displayed in Figure~\ref{DispRel0} change when different values of the parameter $\rho_0$ (\emph{i.e.} the initial ECM density) are considered. Considerations similar to those previously made about the dispersion relations obtained for increasing values of the parameter $\alpha$ apply to the case where increasing values of the parameter $\rho_0$ are considered. In addition to these considerations, the plots in Figure~\ref{DispRelp0} indicate that smaller values of the parameter $\rho_0$, specifically $\rho_0<1$, correlate with a shift in mode selection toward lower modes ({\it cf.} the plots for the linear viscous model~\eqref{ce:d}, the Maxwell model~\eqref{ce:m} and the Jeffrey model~\eqref{ce:3p} in Figure~\ref{DispRelp0}).}

\newpage

\clearpage

\section{Numerical simulations {of a one-dimensional mechanical model of pattern formation}}\label{sec:numerical}
In this section, we verify key results of LSA presented in Section~\ref{sec:lsa} by solving numerically the system of PDEs~\eqref{eq:n}, \eqref{eq:rho} and \eqref{eq:fb} subject to boundary conditions \eqref{bc_1d}. In particular, we report on numerical solutions obtained in the case where equation~\eqref{eq:fb} is complemented with the Kelvin-Voigt model~\eqref{ce:kv} or the Maxwell model~\eqref{ce:m}. A detailed description of the numerical schemes employed is provided in {the Supplementary Material (see `Supplementary Information' document)}.

\paragraph{Set-up of numerical simulations.} We carry out numerical simulations using the parameter values given by~\eqref{eq:param1} and~\eqref{eq:param2}. We choose the endpoints of the spatial domain to be $\ell={0}$ and $L={1}$, and the final time $T$ is chosen sufficiently large so that distinct spatial patterns can be observed at the end of simulations. We consider the initial conditions
\begin{equation}
\label{ic:rand}
n(0,x) = 1 + 0.01\,\epsilon(x) \,, \quad \rho(0,x)  \equiv {\rho_0}  \,, \quad u(0,x) \equiv 0 \,,
\end{equation}
where $\epsilon(x)$ is a normally distributed random variable with mean $0$ and variance $1$ for every $x \in [{0,1}]$. Initial conditions~\eqref{ic:rand} model a scenario where random small perturbations are superimposed to the cell density corresponding to the homogeneous steady state of components $\overline{n}=1$, $\overline{\rho}={\rho_0}$ and $\overline{u}=0$. This is the steady state considered in the LSA undertaken in Section~\ref{sec:LSA}. {Consistent initial conditions for $\dt n(0,x)$, $\dt \rho(0,x)$ and $\dt u(0,x)$ are computed numerically -- details provided in the Supplementary Material (see `Supplementary Information' document).} Numerical computations are performed in MATLAB. 

\begin{figure}[h!]
 \centering
  \includegraphics[width=0.75\linewidth]{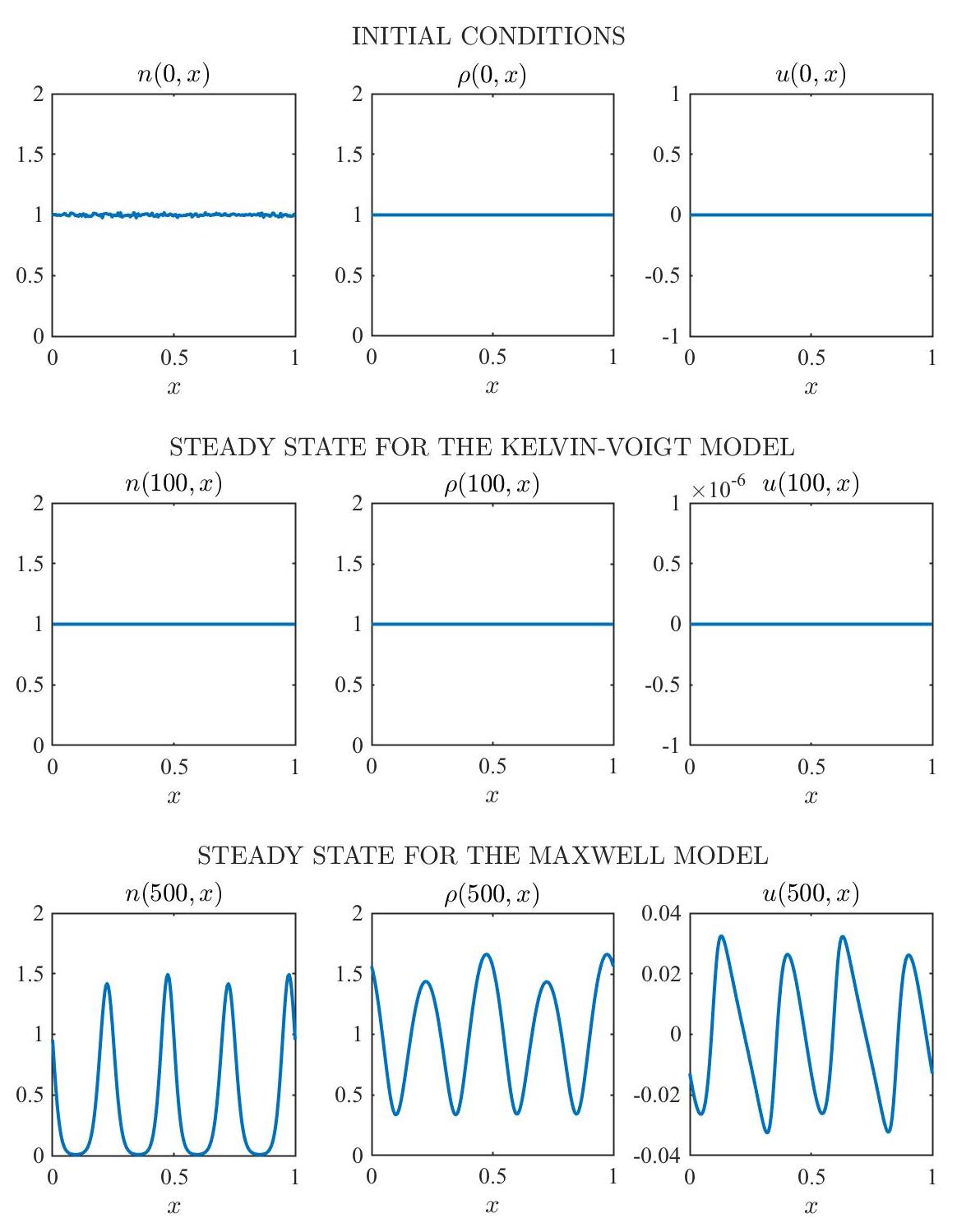}
\caption{{{\bf Simulation results for the Kelvin-Voigt model~\eqref{ce:kv} and the Maxwell model~\eqref{ce:m} under initial conditions~\eqref{ic:rand}.} Cell density $n(t,x)$ (left), ECM density $\rho(t,x)$ (centre) and cell-ECM displacement $u(t,x)$ (right) at $t=0$ (first row) and at steady state obtained solving numerically the system of PDEs~\eqref{eq:n}, \eqref{eq:rho} and \eqref{eq:fb} complemented with the Kelvin-Voigt model~\eqref{ce:kv} (second row) and with the Maxwell model~\eqref{ce:m} (third row), respectively, subject to boundary conditions~\eqref{bc_1d} and initial conditions~\eqref{ic:rand}, for the parameter values given by~\eqref{eq:param1} and~\eqref{eq:param2}.}}
\label{sim1D}
\end{figure}

\paragraph{Main results.} The results obtained are summarised by the plots in Figure~\ref{sim1D}, {together with the corresponding videos provided as supplementary material. The supplementary video `MovS1' displays the solution of the system of PDEs~\eqref{eq:n}, \eqref{eq:rho} and \eqref{eq:fb} subject to the boundary conditions \eqref{bc_1d} and initial conditions~\eqref{ic:rand} for the Kelvin-Voigt model and the Maxwell model from $t=0$ until a steady state, displayed in Figure~\ref{sim1D}, is reached. The supplementary videos `MovS2', `MovS3' and `MovS4' display the solution of the same system of PDEs for the Maxwell model under alternative initial perturbations in the cell density, \emph{i.e.} randomly distributed (`MovS2'), periodic ('MovS3') or randomly perturbed periodic (`MovS4') initial perturbations.\\
The results in Figure~\ref{sim1D} and the supplementary video `MovS1'} demonstrate that, in agreement with the dispersion relations displayed in Figure~\ref{DispRel0}, for the parameter values given by~\eqref{eq:param1} and~\eqref{eq:param2}, small {randomly distributed} perturbations present in the initial cell density: 
\begin{itemize}
\item[-] vanish in the case of the Kelvin-Voigt model, thus leading the cell density to relax to the homogeneous steady state $\overline{n}=1$ and attain numerical equilibrium at $t={100}$ while leaving the ECM density unchanged; 
\item[-] grow in the case of the Maxwell model, resulting in the formation of spatial patterns both in the cell density $n$ and in the ECM density $\rho${, which attain numerical equilibrium at $t=500$}. 
\end{itemize}
Notice that the formation of spatial patterns correlates with the growth of the cell-ECM displacement $u$. In fact, the displacement remains close to zero (\emph{i.e.} $\sim O(10^{-{11}})$) for the Kelvin-Voigt model, whereas it grows with time for the Maxwell model. 
{In addition, the steady state obtained for the Maxwell model in Figure~\ref{sim1D}, together with those obtained when considering alternative initial perturbations (see supplementary videos `MovS2', `MovS3' and `MovS4'), demonstrate that, in agreement with the dispersion relation displayed in Figure~\ref{DispRel0} for the Maxwell model, for the parameter values given by~\eqref{eq:param1} and~\eqref{eq:param2}, under small perturbations in the cell density, be they randomly distributed (\textit{cf.} supplementary video `MovS2'), randomly perturbed periodic (\textit{cf.} supplementary video `MovS3') or periodic (\textit{cf.} supplementary video `MovS4'), the fourth mode is the fastest growing one within the range of unstable modes (\textit{cf.} ${\rm Re}(\psi(k^2))>0$ for $k/\pi$ between 2 and 6, with $\max\big({\rm Re}(\psi(k^2))\big)\approx 4$ in Figure~\ref{DispRel0} for the Maxwell model). In addition, the cellular pattern observed at steady state exhibits 4 large and equally spaced peaks independently of the initial perturbation (\textit{cf.} supplementary videos `MovS1', `MovS2', `MovS3' and `MovS4'). Moreover, all the obtained cellular patterns at steady state exhibit the same structure  -- up to a horizontal shift -- consisting of four large peaks, independently of the initial conditions that is used (\textit{cf.} left panel in the bottom row of Figure~\ref{sim1D} and supplementary videos `MovS2', `MovS3' and `MovS4'). This indicates robustness and consistency in the nature of the saturated nonlinear steady state under specific viscoelasticity assumptions and parameter choices.}

\section{Numerical simulations of a two-dimensional mechanical model of pattern formation}\label{sec:2d}
In this section, we complement the results presented in the previous sections with the results of numerical simulations of a two-dimensional mechanical model of pattern formation in biological tissues. In particular, we report on numerical solutions obtained in the case where the two-dimensional analogue of the system of PDEs~\eqref{eq:n}, \eqref{eq:rho} and~\eqref{eq:fb} is complemented with {a two-dimensional version of the one-dimensional Kelvin-Voigt model~\eqref{ce:kv} or a two-dimensional version of the one-dimensional Maxwell model~\eqref{ce:m}}.

\paragraph{A two-dimensional mechanical model of pattern formation.} The mechanical model of pattern formation defined by the system of PDEs~\eqref{eq:n}, \eqref{eq:rho} and \eqref{eq:fbeg} posed on a two-dimensional spatial domain represented by a bounded set $\Omega \subset\mathbb{R}^2$ with smooth boundary $\partial \Omega$ reads as
\begin{equation}
\label{eq:2d}
\begin{cases}
&\dt n  =   {{\rm div}\left[D\, \nabla n \, - n \, \left(\alpha \, \nabla \rho + \dt \bm{u} \right)\right]  } + r \, n(1-n) \,, \\
&\dt \rho  =  - \,{\rm div} (\rho\, \dt \bm{u} )  \,, \\
&{\rm div}(\bm{\sigma}_m + \bm{\sigma}_c) + \rho \bm{F} = 0 \,,
\end{cases}
\end{equation}
with $t \in (0,T]$, $\bm{x}=(x_1,x_2)^\intercal \in\Omega$ and $\bm{u}=(u_1,u_2)^\intercal$. We close the system of PDEs~\eqref{eq:2d} imposing the two-dimensional version of the periodic boundary conditions~\eqref{bc_1d} on $\partial \Omega$. Furthermore, we use the following two-dimensional analogues of definitions~\eqref{eq:sc} and~\eqref{eq:def}
\begin{equation}
\bm{\s}_c := \frac{\tau n}{1+\lambda n^2}{\Big( \rho + \beta {\rm \Delta} \rho \Big)} \, \bm{I}  \quad \text{and} \quad \bm{F} := - s \, \bm{u} \, ,
\label{eq:2d:sc}
\end{equation}
where $\bm{I}$ is the identity tensor. Moreover, in analogy with the one-dimensional case, we define the stress tensor $\bm{\s}_m$ via the two-dimensional constitutive model that is used to represent the stress-strain relation of the ECM. In particular, we consider the following {generic} two-dimensional {constitutive equation
\begin{equation}
\label{2d:ce}
a_1\dt{\bm{\s}} + a_0\bm{\s} = b_1\dt{\bm{\e}} + b_0{\bm{\e}} + c_1\dt{\theta}\bm{I} + c_0{\theta}\bm{I}  \,.
\end{equation}
This constitutive equation, together with the associated parameter choices reported in Table~\ref{tab:2d:parameters}, summarises the two-dimensional version of the one-dimensional Kelvin-Voigt model~\eqref{ce:kv} and the two-dimensional version of the one-dimensional Maxwell model~\eqref{ce:m} that are considered, which are derived in Appendix~\ref{app:sdmod}.}
{\setlength{\extrarowheight}{7pt}
\begin{table}[]
{\caption{Relations between the parameters in the generic two-dimensional stress-strain constitutive equation~\eqref{2d:ce} and those in the two-dimensional constitutive equations for the Kelvin-Voigt model and the Maxwell model.\label{tab:2d:parameters}}}
\centering
\begin{tabular}{|l|c|c|c|c|c|c|}
\hline
\bf{Generic two-dimensional model} & $\bf{a_1}$ & $\bf{a_0}$ & $\bf{b_1}$ & $\bf{b_0}$ & $\bf{c_1}$ & $\bf{c_0}$ \\ [5pt] \hline 
Kelvin-Voigt model & 0 & $\frac{1}{\eta}$ & 1 & $\frac{E'}{\eta}$ & $\nu'$ & $\frac{ E'\nu'}{\eta}$   \\ [5pt] \hline
Maxwell model & $\frac{1}{E'}$ & $\frac{1}{\eta}$ & 1 & 0 & $\nu'$ & 0   \\ [5pt] \hline
\end{tabular}
\end{table}
}
Here, the strain $\bm{\e}(t,\bm{x})$ and the dilation $\theta(t,\bm{x})$ are defined in terms of the displacement $\bm{u}(t,\bm{x})$ as
\begin{equation}
\label{2d:tensors}
{\bm{\e}}= \frac{1}{2}\big( \nabla\bm{u} + \nabla\bm{u}^\intercal \big)  \quad \text{and} \quad \theta = \nabla \cdot \bm{u} \,.
\end{equation}
Notice that both $\bm{\e}$ and $\theta$ reduce to $\e=\dx u$ in the one-dimensional case. {Amongst the parameters in the stress-strain constitutive equation~\eqref{2d:ce} reported in Table~\ref{tab:2d:parameters} for the two-dimensional Kelvin-Voigt and Maxwell models}, $\eta$ is the shear viscosity, 
\begin{equation}
\label{2d:pp}
E':= \dfrac{E}{1+\nu} \quad \text{and} \quad \nu':= \dfrac{\nu}{1-2\nu} \, ,
\end{equation}
where $\nu$ is Poisson's ratio and $E$ is Young's modulus. As clarified in Appendix~\ref{app:sdmod}, the two-dimensional Maxwell model {in the form~\eqref{2d:ce}} holds under the simplifying assumption that the quotient between the {bulk} viscosity and the {shear} viscosity of the ECM is equal to $\nu'$. 

\paragraph{Set-up of numerical simulations.} We solve numerically the system of PDEs~\eqref{eq:2d} subject to the two-dimensional version of the periodic boundary conditions~\eqref{bc_1d} and complemented with~\eqref{eq:2d:sc}-\eqref{2d:pp}. Numerical simulations are carried out using the following parameter values 
\begin{equation}
\label{eq:param12d}
E={1} \,, \quad \eta={1} \,, \quad D={0.01} \, , \quad \nu=0.25 \, ,
\end{equation}
\begin{equation}
\label{eq:param22d}
{ \alpha=0.05 \,, \quad r=1 \,, \quad s=10 \,, \quad \lambda=0.5 \,, \quad \tau=0.2 \, \quad \beta=0.005 \,,}
\end{equation}
which are chosen for illustrative purposes {and are comparable with nondimensional parameter values that can be found in the extant  literature (see Appendix~\ref{appendix:par} for further details).}
We choose $\Omega = [0,1] \times [0,1]$ and the final time $T$ is chosen sufficiently large so that distinct spatial patterns can be observed at the end of simulations. We consider first the following two-dimensional analogue of initial conditions
~\eqref{ic:rand}
\begin{equation}
\label{ic:2drand}
n(0,x_1,x_2) = 1 + 0.01\,\epsilon(x_1,x_2) \,, \quad \rho(0,x_1,x_2) \equiv 1  \,, \quad \bm{u}(0,x_1,x_2) \equiv \bm{0} \,,
\end{equation}
where $\epsilon(x_1,x_2)$ is a normally distributed random variable with mean $0$ and variance $1$ for each $(x_1,x_2) \in [0,1] \times [0,1]$. {Consistent initial conditions for $\dt n(0,x_1,x_2)$, $\dt \rho(0,x_1,x_2)$ and $\dt \bm{u}(0,x_1,x_2)$ are computed numerically, as similarly done in the one-dimensional case, and numerical computations are performed in MATLAB with a numerical scheme analogous to that employed in the one-dimensional case  -- details provided in the Supplementary Material (see `Supplementary Information' document).}

\begin{figure}[htb!]
 \centering
 \includegraphics[width=1\linewidth]{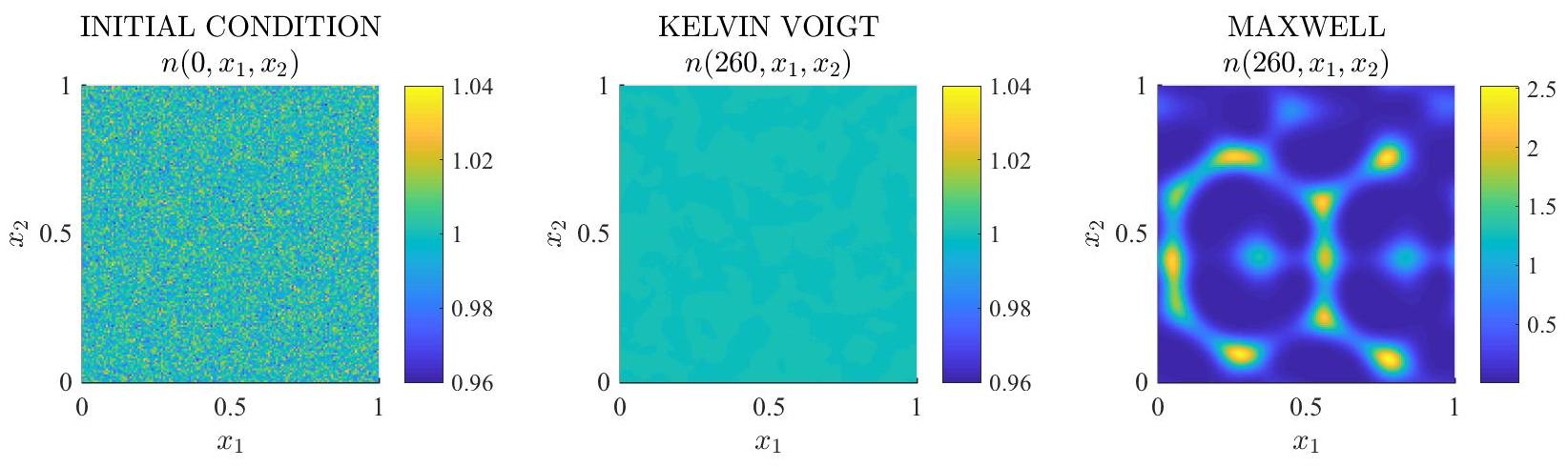}
  \caption{{\bf Simulation results for the two-dimensional {Kelvin-Voigt and} Maxwell models{~\eqref{2d:ce}} under initial conditions{~\eqref{ic:2drand}}.} Cell density $n(t,x_1,x_2)$ at {$t=0$ (left panel) and at $t =260$ for the Kevin-Voigt model (central panel) and the Maxwell model (right panel)} obtained solving numerically the system of PDEs~\eqref{eq:2d} subject to the two-dimensional version of the periodic boundary conditions~\eqref{bc_1d} and initial conditions{~\eqref{ic:2drand}}, complemented with~\eqref{eq:2d:sc}-\eqref{2d:pp}, for the parameter values given by~\eqref{eq:param12d} and~\eqref{eq:param22d}. }
\label{sim2dmain}
\end{figure}

\begin{figure}[htb!]
 \centering
  \includegraphics[width=0.9\linewidth]{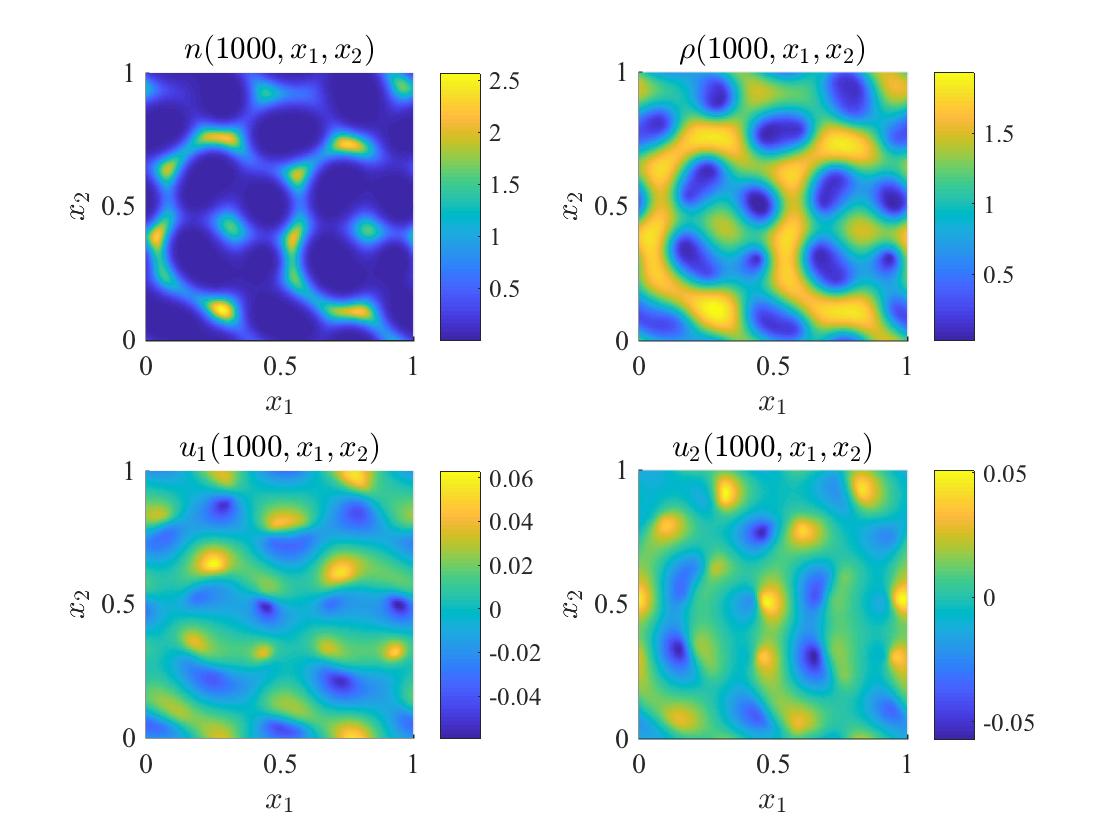}
    \caption{{{\bf Simulation results for the two-dimensional Maxwell model~\eqref{2d:ce} under initial conditions~\eqref{ic:2drand}.} Cell density $n(t,x_1,x_2)$ (top row, left panel), ECM density $\rho(t,x_1,x_2)$ (top row, right panel), first and second components of the cell-ECM displacement ${\bm u}(t,x_1,x_2)$ (bottom row, left panel and right panel, respectively) at $t=1000$
for the Maxwell model obtained solving numerically the system of PDEs~\eqref{eq:2d} subject to the two-dimensional version of the periodic boundary conditions~\eqref{bc_1d} and initial conditions~\eqref{ic:2drand}, complemented with~\eqref{eq:2d:sc}-\eqref{2d:pp}, for the parameter values given by~\eqref{eq:param12d} and~\eqref{eq:param22d}. The random initial perturbation of the cell density is displayed in the left panel of Figure~\ref{sim2dmain}.}}
\label{sim2dMAX}
\end{figure}

\paragraph{Main results.} The results obtained are summarised by the plots in Figures~\ref{sim2dmain} and~\ref{sim2dMAX}, {together with the corresponding videos provided as supplementary material. Solutions of the system of PDEs~\eqref{eq:2d}, together with~\eqref{eq:2d:sc}-\eqref{2d:pp}, subject to initial conditions~\eqref{ic:2drand} and periodic boundary conditions, for the parameter values given by~\eqref{eq:param12d} and~\eqref{eq:param22d}, are calculated both for the Kelvin-Voigt model (see supplementary video `MovS5') and the Maxwell model (see supplementary video `MovS6') according to the parameter changes summarised in Table~\ref{tab:2d:parameters}. The randomly generated initial perturbation in the cell density, together with the cell density at $t=200$ both for the Kelvin-Voigt and the Maxwell model are displayed in Figure~\ref{sim2dmain}, while the solution to the Maxwell model is plotted at a later time in Figure~\ref{sim2dMAX}. Overall, these results demonstrate that, in the scenarios considered here, which are analogous to those considered for the corresponding one-dimensional models, small randomly distributed perturbations present in the initial cell density (\textit{cf.} first panel in Figure~\ref{sim2dmain}): 
\begin{itemize}
\item[-] vanish in the case of the Kelvin-Voigt model, thus leading the cell density to relax to the homogeneous steady state $\overline{n}=1$ and attain numerical equilibrium at $t=260$ (\textit{cf.} second panel of Figure~\ref{sim2dmain}) while leaving the ECM density unchanged (see supplementary video `MovS5'); 
\item[-] grow in the case of the Maxwell model, leading to the formation of spatio-temporal patterns both in the cell density $n$ and in the ECM density $\rho$ (\textit{cf.} third panel of Figure~\ref{sim2dmain}, Figure~\ref{sim2dMAX} and supplementary video `MovS6'), capturing spatio-temporal dynamic heterogeneity arising in the system. 
\end{itemize}
Similarly to the one-dimensional case, the formation of spatial patterns correlates with the growth of the cell-ECM displacement ${\bm u}$. In fact, the displacement remains close to zero (\emph{i.e.} $\sim O(10^{-{11}})$) for the Kelvin-Voigt model (see supplementary video `MovS5'), whereas it grows with time for the Maxwell model (see Figure~\ref{sim2dMAX} and supplementary video `MovS6').
}

\section{Conclusions and research perspectives}\label{sec:discussion}
\paragraph{Conclusions.} We have investigated the pattern formation potential of different stress-strain constitutive equations for the ECM within a one-dimensional mechanical model of pattern formation in biological tissues formulated as the system of implicit PDEs~\eqref{eq:n}, \eqref{eq:rho} and \eqref{eq:fb}. 

The results of linear stability analysis undertaken in Section~\ref{sec:lsa} and the dispersion relations derived therefrom support the idea that {fluid-like} stress-strain constitutive equations ({\it i.e.} the linear viscous model~\eqref{ce:d}, the Maxwell model~\eqref{ce:m} and the {Jeffrey} model~\eqref{ce:3p}) have a pattern formation potential much higher than {solid-like constitutive equations} ({\it i.e.} the linear elastic model~\eqref{ce:s}, the Kelvin-Voigt model~\eqref{ce:kv} and the SLS model~\eqref{ce:sls}).
This is confirmed by the results of numerical simulations presented in Section~\ref{sec:numerical}, which demonstrate that, all else being equal, spatial patterns emerge in the case where the Maxwell model~\eqref{ce:m} is used to represent the stress-strain relation of the ECM, while no patterns are observed when the Kelvin-Voigt model~\eqref{ce:kv} is employed. {In addition, the structure of the spatial patterns presented in Section~\ref{sec:numerical} for the Maxwell model~\eqref{ce:m} is consistent with the fastest growing mode predicted by linear stability analysis.} In Section~\ref{sec:2d}, as an illustrative example, we have also reported on the results of numerical simulations of a two-dimensional version of the model, which is given by the system of PDEs~\eqref{eq:2d} complemented with the two-dimensional {Kelvin-Voigt and} Maxwell models~\eqref{2d:ce}. These results {demonstrate that key features of spatial pattern formation observed in one spatial dimension carry through when two spatial dimensions are considered, thus conferring additional robustness to the conclusions of our work.}

Our findings corroborate the conclusions of~\cite{byrne1996importance} suggesting that prior studies on mechanochemical models of pattern formation relying on the Kelvin-Voigt model of viscoelasticity may have underestimated the pattern formation potential of biological tissues and advocating the need for further empirical work to acquire detailed quantitative information on the mechanical properties of single components of the ECM in different biological tissues, in order to furnish such models with stress-strain constitutive equations for the ECM that provide a more faithful representation of tissue rheology, {\it cf.} \cite{fung1993}.

\paragraph{Research perspectives.}
The dispersion relations given in Section~\ref{sec:lsa} indicate that {there may be parameter regimes whereby solid-like constitutive models of linear viscoelasticity give rise to dispersion relations which exhibit a range of unstable modes, while the dispersion relations obtained using fluid-like constitutive models exhibit singularities, exiting the regime of validity of linear stability analysis.} In this regard, it would be interesting to consider extended versions of the mechanical model of pattern formation defined by the system of PDEs~\eqref{eq:n}, \eqref{eq:rho} and \eqref{eq:fb}, {in order to re-enter the regime of validity of linear stability analysis for the same parameter regimes and verify that in such regimes all constitutive models can produce patterns.} 
For instance, it is known that including long-range effects{, such as long-range diffusion or long-range haptotaxis,} can promote the formation of stable spatial patterns~\cite{moreo2010modelling,oster1983mechanical}, which could be explored through nonlinear stability analysis, as previously done for the case in which the stress-strain relation of the ECM is represented by the Kelvin-Voigt model~\cite{cruywagen1992tissue,lewis1991analysis,maini1988nonlinear}. 
{In particular, weakly nonlinear analysis could provide information on the existence and stability of saturated nonlinear steady states, supercritical bifurcations 
or subcritical bifurcations, which may exist even when the homogeneous steady states are stable to small perturbations according to linear stability analysis~\cite{cross2009pattern}.} 
{Nonlinear analysis would further} enable exploring the existence of possible differences in the spatial patterns obtained when different stress-strain constitutive equations for the ECM are used -- such as amplitude of patterns, perturbation mode selection and geometric structure in two spatial dimensions. {In particular, the base-case dispersion relations given in Section~\ref{sec:lsa} for different fluid-like models of viscoelasticity displayed the same range of unstable modes. This suggests that the investigation of similarities and differences in mode selection between the various models of viscoelasticity could yield interesting results.}
It would also be interesting to construct numerical solutions for the mechanical model defined by the system of PDEs~\eqref{eq:n}, \eqref{eq:rho} and \eqref{eq:fb} complemented with the {Jeffrey} model~\eqref{ce:3p}. For this to be done, suitable extensions of the numerical schemes presented in {the Supplementary Material (see `Supplementary Information' document)} need to be developed. \\
It would also be relevant to systematically assess the pattern formation potential of different constitutive models of viscoelasticity in two spatial dimensions. This would require to relax the simplifying assumption~\eqref{ass:simpl} on the shear and bulk viscosities of the ECM, which we have used to derive the two-dimensional Maxwell model {in the form of~\eqref{2d:ce}}, and, more in general, to find analytically and computationally tractable 
stress-strain-dilation relations, which still remains an open problem~\cite{birman20022d,haghighi2011modeling}. In order to solve this problem, new methods of derivation and parameterisation for constitutive models of viscoelasticity might need to be developed~\cite{valtorta2005dynamic}.\\
As previously mentioned, the values of the model parameters used in this paper have been chosen for illustrative purposes only. Hence, it would be useful to re-compute the dispersion relations and the numerical solutions presented here for  a calibrated version of the model based on real biological data. 
On a related note, {there exists a variety of interesting applications} that could be explored by varying parameter values in the generic constitutive equation~\eqref{eq:fb} {both in space and time. For instance, cell monolayers appear to exhibit solid-like behaviours on small time scales, whereas they exhibit fluid-like behaviours on longer time scales~\cite{tlili2018collective}, and spatio-temporal changes in basement membrane components are known to affect structural properties of tissues during development or ageing, as well as in a number of genetic and autoimmune diseases~\cite{khalilgharibi2021form}. Amongst these, remarkable examples are Alport's syndrome, characterised by changes in collagen IV network due to genetic mutations associated with the disease, diabetes mellitus, whereby high levels of glucose induce significant basement membrane turnover, and cancer. In particular, cancer-associated fibrosis} is a disease characterised by an excessive production of collagen, elastin and proteoglycans, which directly affects the structure of the ECM resulting in alterations of viscoelastic tissue properties~\cite{ebihara2000changes}. Such alterations in the ECM may facilitate tumour invasion and angiogenesis. Considering a calibrated mechanical model of pattern formation in biological tissues, whereby the values of the parameters in the stress-strain constitutive equation for the ECM change during fibrosis progression, may shed new light on the existing connections between structural changes in the ECM components and higher levels of malignancy in cancer~\cite{chandler2019double,park2001lung}. 


 \section*{Acknowledgements}
The authors wish to thank the two anonymous Reviewers  for their thoughtful and constructive feedback on the manuscript (submitted to the \textit{Bulletin of Mathematical Biology}). MAJC gratefully acknowledges the support of EPSRC Grant No. EP/S030875/1 (EPSRC SofTMech$^{\wedge}$MP Centre-to-Centre Award).
 
 \section*{Conflict of interest}
 The authors declare that they have no conflict of interest.
 
\appendix

\renewcommand\thefigure{\thesection.\arabic{figure}}    
\setcounter{figure}{0}  

\setcounter{equation}{0}
\renewcommand\theequation{A.\arabic{equation}}

\newpage

\section{Choice of the parameter values for the baseline parameter sets~\eqref{eq:param1}-\eqref{eq:param2} and~\eqref{eq:param12d}-\eqref{eq:param22d} }\label{appendix:par}
In order not to limit the conclusions of our work by selecting a specific biological scenario, we identified possible ranges of values for each parameter of our model on the basis of the existing literature on mechanochemical models of pattern formation and then define our baseline parameter set by selecting values in the middle of such ranges. In the sensitivity analysis presented in Section~\ref{disprelplots}, we then consider the effect of varying the parameter values within an appropriate range. We first consider the parameters appearing in equations~\eqref{eq:n}, \eqref{eq:rho} and \eqref{eq:fb}, as well as in the initial conditions~\eqref{ic:rand}, and then consider additional parameters appearing in the two-dimensional system~\eqref{eq:2d}-\eqref{2d:pp}, and the associated initial conditions~\eqref{ic:2drand}.
\paragraph{Parameters in the balance equation~\eqref{eq:n}} 
Nondimensional parameter values for the cell motility coefficient $D$ in the literature appear as low as $D=10^{-8}$~\cite{gilmore2012mechanochemical}, and as high as $D=10$~\cite{murray1984generation}, but are generally taken in the range $[10^{-5},1]$~\cite{bentil1991pattern,byrne2003modelling,cruywagen1992tissue,ferrenq1997modelling,maini2002mathematical,murray1988mechanochemical,namy2004critical,olsen1995mechanochemical,perelson1986nonlinear}. Hence, we take $D=0.01$ for our baseline parameter set.
The nondimensional haptotactic sensitivity of cells $\alpha$ takes values in the range 
$[10^{-5},5]$~\cite{bentil1991pattern,cruywagen1992tissue,gilmore2012mechanochemical,murray1988mechanochemical,namy2004critical,olsen1995mechanochemical,perelson1986nonlinear}, and we take $\alpha=0.05$ for our baseline parameter set. 
While most authors ignore cell proliferation dynamics, \textit{i.e.} consider $r=0$ \cite{ambrosi2005review,byrne2003modelling,gilmore2012mechanochemical,murray1988mechanochemical,perelson1986nonlinear}, when present, the rate of cell proliferation takes nondimensional value in the range $[0.02,5]$ \cite{cruywagen1992tissue,olsen1995mechanochemical,perelson1986nonlinear}. Hence, we choose $r=1$ for our baseline parameter set. 
\paragraph{Parameters in the balance equation~\eqref{eq:rho}} While no parameters appear in the balance equation~\eqref{eq:rho}, the value of the parameter $\rho_0$ introduced in Section~\ref{sec:lsa} as the spatially homogenous steady state $\bar{\rho}=\rho_0$, and successively specified to be the initial ECM density in~\eqref{ic:rand} for our numerical simulations, stems from neglected terms in equation~\eqref{eq:rho}. With the exception of~\cite{cruywagen1992tissue} and~\cite{maini2002mathematical} who respectively have $\rho_0=100.2$ and $\rho_0=0.1$, this parameter is usually taken to be $\rho_0=1$ in mechanochemical models ignoring additional ECM dynamics~\cite{bentil1991pattern,cruywagen1992tissue,harris1981fibroblast,manoussaki2003mechanochemical,moreo2010modelling,murray1984cell,murray1984generation,olsen1995mechanochemical,oster1983mechanical,perelson1986nonlinear}.This is generally justified by assuming the steady state $\rho_0$ of equation~\eqref{eq:rho} that is introduced by the additional term, say $S(n,\rho)$, is itself used to nondimensionalise $\rho$, before assuming the dynamics modelled by $S(n,\rho)$ to occur on a much slower timescale than convection driven by the cell-ECM displacement, thus  neglecting this term~\cite[p.328]{murray2001mathematical}, resulting in the nondimensional parameter $\hat{\rho}_0 = 1$. Hence, we take $\rho_0=1$.
 \paragraph{Parameters in the force balance equation~\eqref{eq:fb}} 
 The elastic modulus, or Young modulus, $E$ is usually itself used to nondimensionalise the other parameters in the dimensional correspondent of equation~\eqref{eq:fb} and, therefore, does not appear in the nondimensional system~\cite{bentil1991pattern,gilmore2012mechanochemical,murray1984generation,murray1984generation,murray1988mechanochemical,olsen1995mechanochemical,perelson1986nonlinear}. This corresponds to the nondimensional value $E=1$, which is what we take for our baseline parameter set.
 The viscosity coefficient $\eta$ has been taken with nondimensional values in low orders of magnitude, such as $\eta\sim 10^{-3}-10^{-1}$~\cite{bentil1991pattern,cruywagen1992tissue,gilmore2012mechanochemical,perelson1986nonlinear}, as well as in high orders of magnitude, such as $\eta\sim 10^2-10^3$~\cite{gilmore2012mechanochemical,olsen1995mechanochemical}. It is, however, generally taken to be $\eta=1$~\cite{bentil1991pattern,byrne1996importance,cruywagen1992tissue,murray1984generation,murray1988mechanochemical,perelson1986nonlinear}, which is what we choose for our baseline parameter set. When the constitutive model includes two elastic moduli, \textit{i.e.} for the SLS model~\eqref{ce:sls}, or two viscosity coefficients, \textit{i.e.} for the Jeffrey model~\eqref{ce:3p}, we take $E_1=E_2=E/2 = 0.5$ and  $\eta_1=\eta_2=\eta/2 = 0.5$ as done by~\cite{alonso2017mechanochemical}. 
The cell traction parameter $\tau$ takes nondimensional values spanning many orders of magnitude: it can be found as low as $\tau=10^{-5}$~\cite{ferrenq1997modelling} and as high as $\tau= 10$~\cite{bentil1991pattern,cruywagen1992tissue,perelson1986nonlinear}, but it is generally taken to be of order $\tau\sim 1$~\cite{bentil1991pattern,byrne1996importance,gilmore2012mechanochemical,murray1988mechanochemical,perelson1986nonlinear} and many works consider $\tau\sim 10^{-2}-10^{-1}$~\cite{byrne1996importance,ferrenq1997modelling,murray1984generation,olsen1995mechanochemical}. Hence, for our baseline parameter set we choose $\tau=0.2$. 
The cell-cell contact inhibition parameter $\lambda$ generally takes nondimensional values in the range $ [10^{-2},1]$~\cite{bentil1991pattern,byrne1996importance,murray1988mechanochemical,perelson1986nonlinear}, so we choose $\lambda=0.5$ for our baseline parameter set. The long-range cell traction parameter $\beta$, when present, takes nondimensional values in the range $[10^{-3},10^{-2}]$~\cite{bentil1991pattern,cruywagen1992tissue,gilmore2012mechanochemical,moreo2010modelling,murray1988mechanochemical,perelson1986nonlinear}  so we choose $\beta=0.005$ for our baseline parameter set. 
The elasticity of the external elastic substratum $s$, which is sometimes ignored 
or substituted with a viscous drag, has been taken to have nondimensional values as low as $s \in [10^{-1},1]$~\cite{byrne1996importance,murray1984generation,olsen1995mechanochemical} but is generally chosen in the range $[10,400]$~\cite{bentil1991pattern,gilmore2012mechanochemical,murray1988mechanochemical,perelson1986nonlinear}. Hence, we take $s=10$ for our baseline parameter set.
\paragraph{Parameters in the 2D system~\eqref{eq:2d}-\eqref{2d:pp}} For the parameters in the 2D system~\eqref{eq:2d}-\eqref{2d:pp} and initial condition~\eqref{ic:2drand} that also appear in the equations~\eqref{eq:n}, \eqref{eq:rho}, \eqref{eq:fb} and initial conditions~\eqref{ic:rand}, we make use of the same nondimensional values selected in the one-dimensional case (see previous paragraphs). The Poisson ratio $\nu$, which can only take values in the range $[0.1, 9.45]$, has been estimated to be in the range $[0.2,0.3]$ for the biological tissue considered in mechanochemical models in the current literature~\cite{ambrosi2005review,cruywagen1992tissue,manoussaki2003mechanochemical,moreo2010modelling}. Hence, we choose $\nu=0.25$ for our baseline parameter set. This results in $E'=E/(1+\nu)=0.8$ and $\nu'=\nu/(1-2\nu)=0.5$ according to definitions~\eqref{2d:pp}. In addition, under the simplifying assumption~\eqref{ass:simpl} introduced in Appendix~\ref{app:sdmod}, the bulk viscosity takes the value $\mu=\nu'\eta=0.5\eta=0.5$, which is in agreement with the fact that the bulk and shear viscosities are usually assumed to take values of a similar order of magnitude in the extant literature~\cite{ambrosi2005review,manoussaki2003mechanochemical,moreo2010modelling,murray2003mechanochemical}.

\section{Derivation of the two-dimensional {Kelvin-Voigt and Maxwell models~\eqref{2d:ce}}}
\label{app:sdmod}
Landau \& Lifshitz derived from first principles the stress-strain relations that give the two-dimensional versions of the linear elastic model~\eqref{ce:s} and of the linear viscous model~\eqref{ce:d} in isotropic materials~\cite{landau1970theory}, which read, respectively, as
\begin{equation}
\label{2d:elasticviscous}
{\bm{\s}}_{e} = \frac{E}{1+\nu}\Big( {\bm{\e}}_e + \frac{\nu}{1-2\nu} \theta_e {\bm{I}}\Big) \quad \text{and} \quad {\bm{\s}}_{v} = \eta \, \dt{{\bm{\e}}}_v + \mu \, \dt{\theta}_v {\bm{I}} \, .
\end{equation}
Here, $E$ is Young's modulus, $\nu$ is Poisson's ratio, $\bm{I}$ is the identity tensor, $\eta$ is the shear viscosity and $\mu$ is the bulk viscosity. Moreover, $\bm{\e}_e$ and $\theta_e$ are the strain and dilation under a purely elastic deformation $\bm{u}_e$ while $\bm{\e}_v$ and $\theta_v$ are the strain and dilation under a purely viscous deformation $\bm{u}_v$, which are all defined via~\eqref{2d:tensors}.

In the case of a linearly viscoelastic material satisfying Kelvin-Voigt model, the two dimensional analogue of~\eqref{ce:kv} is simply given by 
\begin{equation}
\label{2d:KV}
{\bm{\s}} = {\bm{\s}}_{e} + {\bm{\s}}_{v} { \,= E'{\bm{\e}} + E'\nu'\theta {\bm{I}} +   \eta \, \dt{{\bm{\e}}} + \mu \, \dt{\theta} {\bm{I}}  \, .}
\end{equation}
{Here $E'$ and $\nu'$ are defined via~\eqref{2d:pp} and there is no distinction between the strain or dilation associated with each component (\textit{i.e.} $\bm{\e}=\bm{\e}_e=\bm{\e}_v$ and $\theta=\theta_e=\theta_v$), as the viscous and elastic components are connected in parallel.} This is the stress-strain constitutive equation that is typically used to describe the contribution to the stress of the cell-ECM system coming from the ECM in two-dimensional mechanochemical models of pattern formation ~\cite{cruywagen1992tissue,ferrenq1997modelling,javierre2009numerical,maini1988nonlinear,manoussaki2003mechanochemical,murray2001mathematical,murray1988mechanochemical,murray1984cell,murray1984generation,murray1983mechanical,olsen1995mechanochemical,oster1983mechanical,perelson1986nonlinear}. 

On the other hand, deriving the two-dimensional analogues of Maxwell model~\eqref{ce:m}, of the SLS model~\eqref{ce:sls} and of the {Jeffrey} model~\eqref{ce:3p} is more complicated due to the presence of elements connected in series. In the case of Maxwell model, using the fact that the overall strain and dilation will be distributed over the different components (\textit{i.e.} $\bm{\e}=\bm{\e}_e+\bm{\e}_v$ and $\theta=\theta_e+\theta_v$) along with the fact that the stress on each component will be the same as the overall stress (\textit{i.e.} $\bm{\sigma}=\bm{\sigma}_e=\bm{\sigma}_v$), one finds
\begin{equation}
\label{2d:Maxwell1}
\frac{1}{\eta}\bm{\s} +\frac{1}{E'}\dt{\bm{\s}} = \dt{\bm{\e}} +\nu'\dt{\theta}\bm{I} + \left(\frac{\mu}{\eta}- \nu' \right)\dt{\theta_v} \, \bf{I} \,,
\end{equation}
with $E'$ and $\nu'$ being defined via~\eqref{2d:pp}. Under the simplifying assumption that
\begin{equation}
\label{ass:simpl}
\dfrac{\mu}{\eta} = \nu'
\end{equation}
the stress-strain constitutive equation~\eqref{2d:Maxwell1} {can be rewritten in the form given by the generic two-dimensional constitutive equation~\eqref{2d:ce} under the parameter choices reported in Table~\ref{tab:2d:parameters}. Dividing~\eqref{2d:KV} by $\eta$, under the simplifying assumption~\eqref{ass:simpl}, the stress-strain constitutive equation for the Kelvin-Voigt model~\eqref{2d:KV} can be rewritten as
$$
\frac{1}{\eta}{\bm{\s}} = \frac{E'}{\eta}{\bm{\e}} + \frac{E'\nu'}{\eta}\theta {\bm{I}} +  \dt{{\bm{\e}}} + \nu' \, \dt{\theta} {\bm{I}}  \, ,
$$
which is in the form given by the generic two-dimensional constitutive equation~\eqref{2d:ce} under the parameter choices reported in Table~\ref{tab:2d:parameters}.}

\bibliographystyle{siam}      
\bibliography{ArXiv_CMAT0324}   

\end{document}


\title{Mechanical models of pattern and form in biological tissues:
the role of stress-strain constitutive equations 
\\
Supplementary Information: Numerical Schemes}
\date{}
\maketitle

In this document (supplementary information) we report on details of the numerical schemes employed to obtain the numerical results presented in the above named publication (main publication). For details of the equations referred to in this document, please see the main publication.
Numerical solutions are computed using \Matlab{}, and the files containing the code corresponding to the schemes presented below can be found on GitLab (\url{https://git-ce.rwth-aachen.de/alf.gerisch/VillaEtAl2021BullMathBiol}).

\setcounter{equation}{0}
\renewcommand\theequation{S.\arabic{equation}}

\section*{Numerical schemes for the system of PDEs~(10), (11) and (16)}\label{appendix:num}
\newcommand{\Dx}{\Delta x}
\newcommand{\Dt}{\Delta t}
\newcommand{\Tau}{\mathrm{T}}
\newcommand{\x}{{\mkern-0mu\times\mkern-0mu}}

Numerical solutions for the system of implicit, time-dependent and spatially one-dimensional PDEs~(10), (11) and (16) are obtained exploiting the Method of Lines. We make use of a uniform discretisation of the spatial domain $[l,L]$ consisting of $\Nx{}+1$ grid points, or grid cell centres, while, at first, leaving the time variable continuous.
We denote the spatial grid width by $\Dx$.
 The normalised cell density $n(t,x)$, the normalised ECM density $\rho(t,x)$ and the displacement of a material point of the cell-ECM system $u(t,x)$ are approximated as
$$
n(t,x_i) \approx N_i(t) \, , \quad \rho(t,x_i) \approx P_i(t) \, , \quad u(t,x_i) \approx U_i(t) \quad \text{for }  i = 0, \ldots, \Nx{} \, .
$$
Thanks to the periodic boundary conditions we have
$$
N_0(t) = N_{\Nx{}}(t) \,, \quad P_0(t) = P_{\Nx{}}(t)  \quad \text{and} \quad U_0(t) = U_{\Nx{}}(t) \,,
$$
and consequently have $3\times \Nx{}$ time-continuous approximations to determine. We collect them in the vectors $N(t)$, $P(t)$, $U(t)$ and denote their time-derivatives by $N'(t)$, $P'(t)$, $U'(t)$.
The discretization of the spatial derivatives in the PDE system will then result in an implicit system of $3\times \Nx{}$ ODEs for the variables $N(t)$, $P(t)$, $U(t)$ and their time-derivatives of the following form
\begin{equation}
\label{eq:MOL}
F(N,P,U,N',P',U') = 
\begin{bmatrix}
f_n(N,P,N',U') \\[5pt]
f_\rho(P,P',U') \\[5pt]
f_u(N,P,U,N',P',U') 
\end{bmatrix} = 0 \;.
\end{equation}
In this system $f_n(N,P,N',U')=0$, $f_\rho(P,P',U')=0$ and $f_u(N,P,U,N',P',U')=0$ are each systems of $\Nx{}$ ODEs obtained, respectively, from PDEs~(10), (11) and (16), using second-order central finite difference approximations for the spatial derivatives and the first-order upwind scheme for the advection terms, as detailed below for each equation. 

In order to solve system~\eqref{eq:MOL}, we make use of the \Matlab{} solver \texttt{ode15i}, which uses a variable-order (orders 1 to 5) backward difference formula (BDF) method in a form suitable to an implicit system of ODEs. Initial conditions $N(0)$, $P(0)$ and $U(0)$ are given by the appropriate equivalent of initial conditions~(28), and we make use of the \Matlab{} function \texttt{decic} to obtain consistent initial conditions $N'(0)$, $P'(0)$ and $U'(0)$ such that~\eqref{eq:MOL} is satisfied at initial time $t=0$.

\paragraph{Useful matrices.} In order to apply the first-order upwind scheme we need to compute variables and derivatives at the grid cell interfaces, \emph{i.e.} half-way between grid points, in addition to those at the grid cell centres. We here clarify the notation adopted throughout the rest of this document. 

The $\Nx{}\x \Nx{}$ matrices ${\bf{M}}_{x}$ and ${\bf{M}}_{xx}$ are used to approximate, using second-order finite differences, the first-order and the second-order derivatives in space, respectively, of a periodic grid function at the grid cell centres and are therefore given by
\begin{equation}
\label{def:Mxx}
{\bf{M}}_{x} := \frac{1}{2\Dx}\,
\begin{bmatrix}
0 & 1 & & & -1 \\[3pt]
-1 & 0 & 1 & &\\
 &  \ddots & \ddots & \ddots  &    \\[3pt]
 & & -1 & 0 & 1 \\[3pt]
 1 & & & -1 & 0
\end{bmatrix}
\qquad \text{and} \qquad
{\bf{M}}_{xx} := \frac{1}{\Dx^2}\,
\begin{bmatrix}
-2 & 1 & & & 1 \\[3pt]
1 & -2 & 1 & &\\
 &  \ddots & \ddots & \ddots  &    \\[3pt]
 & & 1 & -2 & 1 \\[3pt]
 1 & & & 1 & -2
\end{bmatrix} \;.
\end{equation}

We make use of the notation $\overrightharpoonup{\cdot}$ to indicate a shift from the grid cell centres to the (right) grid cell interfaces. In particular to approximate the value of a periodic grid function at these grid cell interfaces we multiply it by the $\Nx{}\x \Nx{}$ matrix
\begin{equation}
\label{def:Mright}
\overrightharpoonup{{\bf{M}}} := \frac{1}{2}\,
\begin{bmatrix}
1 & 1 & & &  \\[3pt]
 & 1 & 1 & &\\
 &  & \ddots & \ddots  &    \\[3pt]
 & &  & 1 & 1 \\[3pt]
 1 & & &  & 1
\end{bmatrix} \,.
\end{equation}
In addition, the $\Nx{}\x \Nx{}$ matrices $\overrightharpoonup{{\bf{M}}}_x$ and $\overleftharpoonup{{\bf{M}}}_x$ are used to approximate the first-order derivatives in space of a periodic grid function at the (right) grid cell interfaces, when the grid function is given in the grid cell centres, and at the grid cell centres, when the grid function is given in the (right) grid cell interfaces, respectively. These are given by
\begin{equation}
\label{def:Mxright}
\overrightharpoonup{{\bf{M}}}_x := \frac{1}{\Dx}\,
\begin{bmatrix}
-1 & 1 & & &  \\[3pt]
 & -1 & 1 & &\\
 &  & \ddots & \ddots  &    \\[3pt]
 & &  & -1 & 1 \\[3pt]
 1 & & &  & -1
\end{bmatrix} 
\qquad \text{and} \qquad
\overleftharpoonup{{\bf{M}}}_x := \frac{1}{\Dx}\,
\begin{bmatrix}
1 &  & & & -1 \\[3pt]
-1 & 1 &  & &\\
 &  \ddots & \ddots &   &    \\[3pt]
 & & -1 & 1 &  \\[3pt]
  & & & -1 & 1
\end{bmatrix} \;.
\end{equation}
Note that, even though these two matrices are multiplied by $1/\Dx$, they still stem from second-order finite difference approximations, calculated on a staggered grid shifted by half the grid cell width.

\textbf{Convention:} In the formulas which follow below, we use the convention that any product of a matrix from above with a vector of length $\Nx{}$ is a matrix-vector product but any operation between two vectors, in particular multiplication, division, or exponentiation, are understood element-wise.

\paragraph{Numerical scheme for the balance equation~(10).} We rewrite the balance equation~(10) as
$$
\dt n \,-\, D \dxx n \,+\, \dx( \phi\, n) \,-\, rn(1-n) \,=\,0 \quad \text{with} \quad \phi =  \alpha\, \dx\rho + \dt u \, ,
$$
which, upon spatial discretisation, leads to the following system of $\Nx{}$ ODEs
\begin{equation}
\label{eq:ndiscnum}
f_n(N,P,N',U') = N' - D \,{\bf{M}}_{xx}\,N + \mathrm{A} \big(\overrightharpoonup{\Phi}, N\big) - rN(1-N) = 0
\end{equation}
with $\overrightharpoonup{\Phi}$ indicating the advective velocity computed at the grid cell interfaces, that is
\begin{equation}
\label{eq:phidiscnum}
\overrightharpoonup{\Phi} = \alpha \,{\overrightharpoonup{\bf{M}}}_{x}\,P + {\overrightharpoonup{\bf{M}}} \,U'\, ,
\end{equation}
and the matrices ${\bf{M}}_{xx}$, ${\overrightharpoonup{\bf{M}}}$ and ${\overrightharpoonup{\bf{M}}}_{x}$ are defined in~\eqref{def:Mxx}, \eqref{def:Mright} and~\eqref{def:Mxright}, respectively. 
The function $\mathrm{A} \big(\overrightharpoonup{\Phi}, N\big)$ computes the contribution of advection, given advective velocity and advected quantity as inputs, at the grid cell centres as
\begin{equation}
\label{def:Aflux}
\mathrm{A} \big(\overrightharpoonup{\Phi}, N\big) := \overleftharpoonup{{\bf{M}}}_x \, \overrightharpoonup{\mathcal{F}}\big(\overrightharpoonup{\Phi}, N\big) \,
\end{equation}
where the matrix $\overleftharpoonup{{\bf{M}}}_x$ is defined in~\eqref{def:Mxright} and the advective flux $\overrightharpoonup{\mathcal{F}}$ at the grid cell interfaces is computed using first-order upwinding, \emph{i.e.}
\begin{equation}
\label{def:flux}
\Big[ \overrightharpoonup{\mathcal{F}} \big(\overrightharpoonup{\Phi}, N \big) \Big]_i :=\,
\begin{cases}
\big(\overrightharpoonup{\Phi}_i\big)^+\,N_i + \big(\overrightharpoonup{\Phi}_i\big)^-\,N_{i+1}  \quad\; &\text{for} \;\; i=1,...,\Nx{}-1 \\[5pt]
\big(\overrightharpoonup{\Phi}_{\Nx{}}\big)^+\,N_{\Nx{}} + \big(\overrightharpoonup{\Phi}_{\Nx{}}\big)^-\,N_{1}  \qquad &\text{for} \;\; i=\Nx{}
\end{cases} 
\end{equation}
with $(\cdot)^+$ and $(\cdot)^-$ being the positive and negative parts of the input variable, \emph{i.e.}
\begin{equation}
\label{def:pn}
(\Phi)^+ := \max(0,\Phi) \qquad \text{and} \qquad (\Phi)^- := \min(0,\Phi) \;.
\end{equation}

\paragraph{Numerical scheme for the transport equation~(11).} We rewrite the transport equation~(11) as
$$
\dt \rho \,+\, \dx( \dt u\, \rho) \,=\, 0
$$
which, upon spatial discretisation, leads to the following system of $\Nx{}$ ODEs
\begin{equation}
\label{eq:rhodiscnum}
f_\rho(P,P',U') = P' +  \mathrm{A} \big(\overrightharpoonup{{\bf{M}}}\,U', P\big) \, = 0 \;,
\end{equation}
where the function $\mathrm{A} \big(\overrightharpoonup{{\bf{M}}}\,U', P\big)$ is defined in~\eqref{def:Aflux}, together with definitions~\eqref{def:flux} and~\eqref{def:pn}, with advection velocity given by $U'$ calculated at the cell interfaces using $\overrightharpoonup{{\bf{M}}}$ defined in~\eqref{def:Mright}.

\paragraph{Numerical scheme for the force-balance equation~(16).} We solve the system of PDEs~(10), (11) and (16) for the Kelvin-Voigt~(3) and the Maxwell~(4) models. In these cases we have  $b_2=a_2=0$, and the force-balance equation~(16) reads as
$$
b_1 \,\partial^3_{xxt} u +b_0 \,\dxx u \,-\, a_1 s\, \dt (\rho u) \,-\,  a_0\, s\rho u \,+\,  \dx \big( a_1\,\partial_{t} \s_c + a_0\,\s_c \big) = 0 \quad \text{with} \quad \s_c = \tau \, \dfrac{n}{1+\lambda \, n^2}  \big(\rho + \beta\, \partial^2_{xx}\rho \big) \,.
$$
Upon spatial discretisation, this leads to the following system of $\Nx{}$ ODEs
\begin{equation}
\label{eq:udiscnum}
f_u(N,P,U,N',P',U')  = b_1 {\bf{M}}_{xx} U' + b_0 {\bf{M}}_{xx} U - a_1s(PU)' - a_0sPU + {\bf{M}}_{x} \Tau_1(N,P,N',P') = 0
\end{equation}
with
\begin{equation}
\label{eq:udiscnumT}
\Tau_1(N,P,N',P') =  \;\tau \big[\, a_1 \, \Lambda_2(N) N' {\bf{M}}_{T1}P  + a_1 \,  \Lambda_1(N){\bf{M}}_{T1}P' + a_0 \, \Lambda_1(N) {\bf{M}}_{T1}P\,\big]  \,,
\end{equation}
where the functions $\Lambda_1$ and $\Lambda_2$ are defined as
\begin{equation}
\label{def:Lambda12}
\Lambda_1(N):= \dfrac{N}{1+\lambda\,N^2}  \qquad \text{and its derivative} \qquad \Lambda_2(N):= \dfrac{1-\lambda\,N^2}{(1+\lambda\,N^2)^2} \,, 
\end{equation}
while the $\Nx{}\x \Nx{}$ matrix ${\bf{M}}_{T1}$ is given by
\begin{equation}
\label{def:MT}
{\bf{M}}_{T1}:= {\bf{I}} + \beta \,{\bf{M}}_{xx} \,,
\end{equation}
where ${\bf{I}}$ is the $\Nx{}\x \Nx{}$ identity matrix and ${\bf{M}}_{x}$ is defined in~\eqref{def:Mxx}. \\
This scheme is valid as long as $b_2=a_2=0$ and can therefore also be applied when considering the linear elastic model~(1), the linear viscous model~(2), and the SLS model~(5). On the other hand, in the case where $b_2 \neq 0$ (\emph{i.e.} when the Jeffrey model~(6) is considered) the above numerical scheme cannot be directly employed due to the presence of a second-order derivative in $t$. We could still, however, take a similar approach and make use of the \texttt{ode15i} solver by introducing extra variables for the first-order derivatives in $t$ of $n$ and $\rho$, thus formally reducing the PDE~(16) to first-order in time, at the cost of increasing the number of equations in the Method of Lines ODE system.

\section*{Numerical scheme for the system of PDEs~(29)}
Similarly as done for the spatially one-dimensional model, numerical solutions for the system of implicit, time-dependent and spatially two-dimensional PDEs~(29), together with~(30)-(32), are obtained exploiting the Method of Lines. We make use of a uniform discretisation of the square spatial domain $[l,L]\times[l,L]$ consisting of $(\Nx{}+1)\x(\Nx{}+1)$ grid points, while leaving the time variable continuous.
The spatial grid width, in both spatial directions, is denoted by $\Dx$ again.
The normalised cell density $n(t,x_1,x_2)$, the normalised ECM density $\rho(t,x_1,x_2)$ and the displacement of a material point of the cell-ECM system ${\bm u}(t,x_1,x_2)=\big( u_1(t,x_1,x_2),u_2(t,x_1,x_2)\big)^{\intercal}$ are approximated as 
$$
n(t,x_i,x_j) \approx {N}_{i,j}(t) \, , \quad \rho(t,x_i,x_j) \approx {P}_{i,j}(t) \quad \text{for }  i,j = 0, \ldots, \Nx{} \, ,
$$
$$
u_1(t,x_i,x_j) \approx ({U}_1)_{i,j}(t)\, , \quad u_2(t,x_i,x_j) \approx ({U}_2)_{i,j}(t) \quad \text{for }  i,j = 0, \ldots, \Nx{} \, .
$$
Thanks to the periodic boundary conditions, we can drop the index values $i=0$ and $j=0$ and consequently have $4\times \Nx{}^2$ time-continuous approximations to determine. 
We collect them in the matrices $N(t)$, $P(t)$, $U_1(t)$, $U_2(t)$ and denote their time-derivatives by $N'(t)$, $P'(t)$, $U_1'(t)$, $U_2'(t)$.
The discretization of the spatial derivatives in the PDE system will then result in an implicit system of $4\times \Nx{}^2$ ODEs for the variables $N(t)$, $P(t)$, $U_1(t)$, $U_2(t)$ and their time-derivatives of the following form
\begin{equation}
\label{eq:MOL2d}
F(N,P,U_1,U_2,N',P',U'_1,U'_2) = 
\begin{bmatrix}
f_n(N,P,N',U'_1,U'_2) \\[5pt]
f_\rho(P,P',U'_1,U'_2) \\[5pt]
f_{u_1}(N,P,U_1,U_2,N',P',U'_1,U'_2) \\[5pt]
f_{u_2}(N,P,U_1,U_2,N',P',U'_1,U'_2) 
\end{bmatrix} = 0 \;.
\end{equation}
In this system
$f_n(N,P,N',U'_1,U'_2)=0$, $f_\rho(N,P,N',U'_1,U'_2)=0$, $f_{u_1}(N,P,U_1,U_2,N',P',U'_1,U'_2)=0$ and
$f_{u_2}(N,P,U_1,U_2,N',P',U'_1,U'_2)=0$ are each systems of $\Nx{}^2$ ODEs obtained from the system of PDEs~(29), using second-order central finite difference approximations for the spatial derivatives and the first-order upwind scheme for the advection terms, as detailed below for each equation. 

In order to solve system~\eqref{eq:MOL2d}, we make, similarly to the spatially one-dimensional case, use of the \Matlab{} solver \texttt{ode15i}. Initial conditions $N(0)$, $P(0)$, $U_1(0)$ and $U_2(0)$ are given by the appropriate equivalent of initial conditions~(36), and we make use of the \Matlab{} function \texttt{decic} to obtain consistent initial conditions $N'(0)$, $P'(0)$, $U'_1(0)$ and $U'_2(0)$ such that~\eqref{eq:MOL2d} is satisfied at initial time $t=0$.

\paragraph{Useful functions} In order to solve the system~\eqref{eq:MOL2d} we need to compute variables and derivatives at the grid cell centers and interfaces, both in the $x_1$- and the $x_2$-direction. We here introduce the functions that will be used in the rest of this document to compute the aforementioned quantities in the different directions. These rely on the fact that the matrices~\eqref{def:Mxx}-\eqref{def:Mxright} act on column vectors and therefore, when applied to an $\Nx{}\x \Nx{}$ argument matrix, they will act on each column of that, which in our framework corresponds to computing the quantity of interest in the $x_1$-direction. In order to compute the same quantities in the $x_2$-direction, we need the operating matrix to act on each row of the argument matrix of interest, which can be achieved by matrix transposition of the argument matrix before and of the product matrix after matrix multiplication. Hence the functions ${\bf{M}}_{x1}(N)$ and ${\bf{M}}_{x2}(N)$ are used to approximate the first-order derivative of the variable of interest, say $N$, at the grid cell centres in the $x_1$- and $x_2$-directions respectively, and are defined as
\begin{equation}
\label{def:Mx1}
{\bf{M}}_{x1}(N):= \, {\bf{M}}_{x}\, N \,, \qquad \text{and} \qquad {\bf{M}}_{x2}(N):= \, \big[ {\bf{M}}_{x}\, N^\intercal \big]^\intercal \,,
\end{equation}
where the matrix ${\bf{M}}_{x}$ is defined in~\eqref{def:Mxx}. Similarly, the functions ${\bf{M}}_{xx1}(N)$ and ${\bf{M}}_{xx2}(N)$ are used to approximate the second-order derivative of the variable of interest at the grid cell centres in the $x_1$- and $x_2$-directions, respectively, and are defined as 
\begin{equation}
\label{def:Mxx1}
{\bf{M}}_{xx1}(N):= \, {\bf{M}}_{xx}\, N \,, \qquad \text{and} \qquad {\bf{M}}_{xx2}(N):= \, \big[ {\bf{M}}_{xx}\, N^\intercal \big]^\intercal \,,
\end{equation}
where the matrix ${\bf{M}}_{xx}$ is defined in~\eqref{def:Mxx}. Then the function ${\bf{M}}_{x1x2}(N)$ is used to approximate the second-order mixed derivative in space at the grid cell centres and is defined as
\begin{equation}
\label{def:Mx1x2}
{\bf{M}}_{x1x2}(N):= \, {\bf{M}}_{x2}\big( {\bf{M}}_{x1}(N)\big) \,= \,    \big[ {\bf{M}}_{x}\, \big( {\bf{M}}_{x}\, N \big)^\intercal \big]^\intercal\,.
\end{equation}
In order to approximate the value of a variable in the centres of the (right or upper) grid cell interfaces in the $x_1$- and $x_2$-direction, we make use of the functions $\overrightharpoonup{{\bf{M}}}_1$ and $\overrightharpoonup{{\bf{M}}}_2$, respectively, which are defined as
\begin{equation}
\label{def:Mright1}
\overrightharpoonup{{\bf{M}}}_1(N):= \, \overrightharpoonup{{\bf{M}}}\, N \,, \qquad \text{and} \qquad \overrightharpoonup{{\bf{M}}}_2(N):= \, \big[ \overrightharpoonup{{\bf{M}}}\, N^\intercal \big]^\intercal \,,
\end{equation}
with the matrix $\overrightharpoonup{{\bf{M}}}$ defined in~\eqref{def:Mright}. In a similar fashion we define the functions $\mathrm{A}_1 \big(\overrightharpoonup{v_1}, N\big)$ and $\mathrm{A}_2\big(\overrightharpoonup{v_2}, N\big)$ which approximate the contribution of advection in the $x_1$- and $x_2$-direction, respectively, given as input the advective velocity at the grid cell interfaces in the direction of interest -- say ${v_1}$ and ${v_2}$ are, respectively, the first and second components of the advective velocity -- and the advected quantity. These are given by
\begin{equation}
\label{def:A1:flux}
\mathrm{A}_1 \big(\overrightharpoonup{v_1}, N\big):=\,\mathrm{A}\big(\overrightharpoonup{v_1}, N\big) \qquad \text{and} \qquad \mathrm{A}_2 \big(\overrightharpoonup{v_2}, N\big) :=\,\big[ \mathrm{A} \big(\overrightharpoonup{v_2}^\intercal, N^\intercal\big)\big]^\intercal \,,
\end{equation}
with the function $\mathrm{A} \big(\overrightharpoonup{v_1}, N\big)$ given by~\eqref{def:Aflux} together with definitions~\eqref{def:flux} and~\eqref{def:pn}.

\textbf{Convention:} With the definitions above, we have hidden all applications of the matrices from the spatially one-dimensional case in newly defined functions. Consequently, in the formulas which follow below, we use the convention that any further operation between matrices, in particular multiplication, division, or exponentiation, are understood element-wise.

\paragraph{Numerical scheme for the balance equation~(29)$_1$.} We rewrite the balance equation~(29)$_1$ as
$$
\dt n \,-\, D \big[ \partial^2_{x_1x_1} n + \partial^2_{x_2x_2} n \big] \,+\, \partial_{x_1}( \phi_1\, n) \,+\, \partial_{x_2}( \phi_2\, n) \,-\, rn(1-n) \,=\,0 \quad \text{with} \quad \phi_i =  \alpha\, \partial_{x_i}\rho + \dt u_i \;\; i=1,2 \, ,
$$
which, upon spatial discretisation, leads to the following system of $\Nx{}^2$ ODEs
\begin{equation}
\label{eq:ndiscnum2d}
f_n(N,P,N',U'_1,U'_1) = N' - D \big[ \,{\bf{M}}_{xx1}(N) +\,{\bf{M}}_{xx2}(N)\big]\, + \mathrm{A}_1 \big(\overrightharpoonup{\Phi}_1, N\big) + \mathrm{A}_2 \big(\overrightharpoonup{\Phi}_2, N\big) - rN(1-N) = 0
\end{equation}
with the functions ${\bf{M}}_{xx1}(\cdot)$ and ${\bf{M}}_{xx2}(\cdot)$ defined in~\eqref{def:Mxx1}, and the components of the advective velocity at the grid cell interfaces given by
\begin{equation}
\label{eq:phidiscnum2d}
\overrightharpoonup{\Phi}_1 = \alpha \,{\overrightharpoonup{\bf{M}}}_{x1}(P) + {\overrightharpoonup{\bf{M}}}_1 (U'_1) \quad \text{and} \quad \overrightharpoonup{\Phi}_2 =\alpha \,{\overrightharpoonup{\bf{M}}}_{x2}(P )+ {\overrightharpoonup{\bf{M}}}_2 (U'_2) \, ,
\end{equation}
where functions ${\overrightharpoonup{\bf{M}}}_{x1}(\cdot)$ and ${\overrightharpoonup{\bf{M}}}_{x2}(\cdot)$ are defined in~\eqref{def:Mx1}, ${\overrightharpoonup{\bf{M}}}_{1}(\cdot)$ and ${\overrightharpoonup{\bf{M}}}_{2}(\cdot)$ are defined in~\eqref{def:Mright1}, and functions $\mathrm{A}_1 (\cdot,\cdot)$ and $\mathrm{A}_2 (\cdot,\cdot)$ are defined in~\eqref{def:A1:flux}.

\paragraph{Numerical scheme for the transport equation~(29)$_2$.} We rewrite the transport equation~(29)$_2$ as
$$
\dt \rho \,+\, \partial_{x_1}( \dt u_1\, \rho) \,+\, \partial_{x_2}( \dt u_2\, \rho) \,=\, 0
$$
which, upon spatial discretisation, leads to the following system of $\Nx{}^2$ ODEs
\begin{equation}
\label{eq:rhodiscnum2d}
f_\rho(P,P',U'_1,U'_2) = P' + \mathrm{A}_1 \big({\overrightharpoonup{\bf{M}}}_1 (U'_1), P\big) + \mathrm{A}_2 \big({\overrightharpoonup{\bf{M}}}_2 (U'_2), P\big)  \, = 0 \;,
\end{equation}
where the functions $\mathrm{A}_1 (\cdot,\cdot)$ and $\mathrm{A}_2 (\cdot,\cdot)$ are defined in~\eqref{def:A1:flux} and functions ${\overrightharpoonup{\bf{M}}}_{1}(\cdot)$ and ${\overrightharpoonup{\bf{M}}}_{2}(\cdot)$ are defined in~\eqref{def:Mright1}.

\newcommand{\daa}{\partial^2_{x_1x_1}}
\newcommand{\dab}{\partial^2_{x_1x_2}}
\newcommand{\dbb}{\partial^2_{x_2x_2}}

\paragraph{Numerical scheme for the force balance equation~(29)$_3$.} We rewrite the first component of the force balance equation~(29)$_3$, complemented with (30)-(32), as
\begin{equation}
\begin{split}
&{b_1 \Big( \daa \partial_t u_1 + \frac{1}{2}\big[\dbb \partial_t u_1 + \dab \partial_t u_2 \big] \Big) +
b_0 \Big( \daa u_1 + \frac{1}{2}\big[\dbb u_1 + \dab u_2 \big] \Big) \,+ } \\[5pt]
&{ c_1 \big( \daa \partial_t u_1 + \dab \partial_t u_2 \big) + c_0 \big( \daa u_1 + \dab u_2 \big) \,+ }\\[8pt]
&{\partial_{x_1}[a_1 \dt \s_c + a_0 \s_c] - a_1 s (u_1 \partial_t\rho + \rho \partial_t u_1) - a_0 s \rho u_1 = 0 \;,}
\end{split}
\end{equation}
and, similarly, we rewrite the second component as
\begin{equation}
\begin{split}
&{b_1 \Big( \dbb \partial_t u_2 + \frac{1}{2}\big[\dab \partial_t u_1 + \daa \partial_t u_2 \big] \Big) +
b_0 \Big( \dbb u_2 + \frac{1}{2}\big[\dab u_1 + \daa u_2 \big] \Big) \,+ } \\[5pt]
&{ c_1 \big( \dbb \partial_t u_2 + \dab \partial_t u_1 \big) + c_0 \big( \dbb u_2 + \dab u_1 \big) \,+ }\\[8pt]
&{\partial_{x_2}[a_1 \dt \s_c + a_0 \s_c] - a_1 s (u_2 \partial_t\rho  + \rho \partial_t u_2) - a_0 s \rho u_2 = 0 \;,}
\end{split}
\end{equation}
where $\s_c$ is defined by
\begin{equation}
\s_c = \tau \frac{n}{1+\lambda n^2} \big( \rho + \beta \daa\rho + \beta \dbb\rho \big) \;.
\end{equation}

\noindent Upon spatial discretisation, these lead to the following systems of $\Nx{}^2$ ODEs
\begin{equation}
\label{eq:udiscnum2d1t}
\begin{split}
f_{u1}(N,P,U,N',P',U')  =&\; b_1 \big(\, {\bf{M}}_{xx1} (U'_1) +\frac{1}{2}(\,{\bf{M}}_{xx2} (U'_1) + {\bf{M}}_{x1x2} (U'_2) \,)\,\big) \,+ \\[5pt]
&b_0 \big(\, {\bf{M}}_{xx1} (U_1) +\frac{1}{2}(\,{\bf{M}}_{xx2} (U_1) + {\bf{M}}_{x1x2} (U_2) \,)\,\big) \,+ \\[5pt]
&\; c_1 \big( \,{\bf{M}}_{xx1} (U'_1)  + {\bf{M}}_{x1x2} (U'_2)\,  \big) + c_0  \big( \,{\bf{M}}_{xx1} (U_1)  + {\bf{M}}_{x1x2} (U_2)  \,\big) \,+ \\[5pt]
&\; {\bf{M}}_{x1} \big( \,\Tau_2(N,P,N',P') \,\big) - a_1s ( PU'_1 + P'U_1 )- a_0sPU_1 \,= \,0 
\end{split}
\end{equation}
and
\begin{equation}
\label{eq:udiscnum2d2t}
\begin{split}
f_{u2}(N,P,U,N',P',U')  =&\; b_1 \big( \,{\bf{M}}_{xx2} (U'_2)+\frac{1}{2}(\,{\bf{M}}_{x1x2} (U'_1) + {\bf{M}}_{xx1} (U'_2)\, )\,\big) \,+ \\[5pt]
&b_0 \big(\, {\bf{M}}_{xx2} (U_2) +\frac{1}{2}(\,{\bf{M}}_{x1x2} (U_1 )+ {\bf{M}}_{xx1} (U_2) \,)\,\big) + \\[5pt]
&\; c_1 \big(\, {\bf{M}}_{xx2} (U'_2)  + {\bf{M}}_{x1x2} (U'_1 ) \,\big) + c_0  \big(\, {\bf{M}}_{xx2} (U_2) + {\bf{M}}_{x1x2} (U_1) \, \big) \,+ \\[5pt]
&\; {\bf{M}}_{x2} \big( \, \Tau_2(N,P,N',P')\,\big) - a_1s(PU'_2 + P'U_2 ) - a_0sPU_2 \,= \,0 \;.
\end{split}
\end{equation}
Here
\begin{equation}
\label{eq:udiscnumT2d}
\Tau_2(N,P,N',P') =  \;\tau \big[\, a_1 \, \Lambda_2(N) N' \,{\bf{M}}_{T2}(P)  + a_1 \,  \Lambda_1(N)\,{\bf{M}}_{T2}(P' )+ a_0 \, \Lambda_1(N) \,{\bf{M}}_{T2}(P)\,\big]  \,,
\end{equation}
where the functions $\Lambda_1$ and $\Lambda_2$ are defined as in~\eqref{def:Lambda12}, while the function ${\bf{M}}_{T2}(P)$ is given by
\begin{equation}
\label{def:MT2d}
{\bf{M}}_{T2}(P):= P + \beta \big[\,{\bf{M}}_{xx1}(P) + {\bf{M}}_{xx2}(P)  \,\big] \,,
\end{equation}
where the functions ${\bf{M}}_{xx1}(\cdot)$ and ${\bf{M}}_{xx2}(\cdot)$ are defined in~\eqref{def:Mxx1}.

\paragraph{Remark:} 
The \Matlab{} solver \texttt{ode15i} allows for the specification of the sparsity pattern of Jacobian matrices. In particular in the spatially two-dimensional simulations this leads, in comparison to not specifying these patterns, to substantial savings in required CPU time. For details on these patterns we refer to the available \Matlab{} implementation for the numerical solution of the PDE systems.